\documentclass[twoside,titlepage]{article}
\usepackage{amsmath, amsthm, amssymb} 
\usepackage{mathrsfs}
\usepackage{amsfonts}
\usepackage{hyperref} 
\usepackage{graphicx} 
\numberwithin{equation}{section} 

\DeclareMathAlphabet{\mathpzc}{OT1}{pzc}{m}{it}
\DeclareMathAlphabet{\mathpzc}{OT1}{pzc}{m}{it}

\begin{document}


\newcommand{\be}{\begin{equation}}
\newcommand{\e}{\end{equation}}
\newcommand{\bea}{\begin{equation*}}
\newcommand{\ea}{\end{equation*}}
\newcommand{\la}{\label}
\newcommand{\bu}{\bullet}

\begin{titlepage}

\today

\begin{center}

\hfill{\tt WIS/11/12-JUNE-DPPA}\\


\vskip 20mm

{\Large{\bf 4d $\mathcal N=2$ superconformal linear quivers with type IIA duals}}

\vskip 10mm

{\bf Ofer~Aharony, Leon~Berdichevsky, and Micha~Berkooz\footnote{Emails: ofer.aharony, leon.berdichevsky, micha.berkooz@weizmann.ac.il}
}

\vskip 4mm
{\em Department of Particle Physics and Astrophysics,}\\
{\em Weizmann Institute of Science,}\\
{\em Rehovot 76100, Israel}\\
[2mm]

\end{center}
\vskip 2cm

\begin{center} {\bf ABSTRACT }\end{center}
\begin{quotation}
\noindent
%
We discuss the gravity duals of 4d $\mathcal N=2$ superconformal field
theories (SCFTs) arising from the low-energy limit of
brane configurations of D4-branes stretched
between and intersecting NS5-branes and D6-branes. This gives
rise to a product of $SU(N_i)$ groups, with bi-fundamental matter
between adjacent groups, and extra fundamental hypermultiplets.
The most general configuration in 11d (or type IIA) supergravity that is dual
to a 4d $\mathcal N=2$ SCFT (when the dual of this SCFT is a weakly
curved background) was written down by Gaiotto
and Maldacena, but finding it explicitly involves solving a complicated
Toda equation. This equation simplifies only when the solution can
be reduced to type IIA supergravity, so we ask for which SCFTs of
this type is there a type IIA dual that is weakly coupled and
weakly curved (away from NS5-branes and D6-branes). We find that such
solutions (a special case of which was analyzed by Reid-Edwards
and Stefanski) exist when there is a large number of gauge
groups, with large ranks, and with large 't Hooft couplings for
all but a finite number of groups. The general solution of this
type is given by solving an axially symmetric Laplace
equation in three dimensions, with specific boundary conditions.
We match the parameters of the 4d SCFTs, including the exactly
marginal coupling constants, with the boundary conditions for
the Laplace equation.

\end{quotation}

\vfill

\end{titlepage}

\tableofcontents





\section{Introduction and summary of results}

Finding the gravitational solutions describing configurations of branes intersecting other branes, or branes ending
on other branes, is a challenging problem, which (even in the classical gravity limit)
only has a solution in some very special cases.
In addition to its intrinsic interest, this problem is also interesting in the context of the
duality between gravitational theories and quantum field theories, since many interesting quantum field
theories arise as the low energy limit of  branes ending on and intersecting other branes (following \cite{Ganor,Hanany:1996ie}). Finding the corresponding gravitational solutions would enable (when they are weakly coupled and curved) studying the corresponding quantum field theories at strong coupling.
In this paper we study this problem for a specific class of 4d ${\cal N}=2$ superconformal field theories (SCFTs).

Recently, the solutions to type IIB supergravity describing the near-horizon limit of D3-branes ending on 5-branes \cite{Aharony:2011yc} and D3-branes stretched between 5-branes \cite{Assel:2011xz} were found, using the general classification of type IIB supergravity solutions with $OSp(4|4)$ symmetry \cite{D'Hoker:2007xy,D'Hoker:2007xz}. The former are dual to the 4d $\mathcal N=4$ supersymmetric Yang-Mills theory on a half line with various boundary conditions that preserve 16 supercharges, while the latter are dual to the 3d $\mathcal N=4$ SCFTs analyzed by Gaiotto and Witten \cite{Gaiotto:2008ak}. The gravity solutions include the D5-branes and NS5-branes that were present in the original brane configuration; under suitable conditions, the solutions are weakly coupled and weakly curved away from the brane sources. Following these results, it is natural to search for gravity solutions describing configurations of branes intersecting and ending on other branes in type IIA string theory, that preserve the same amount of supersymmetry.

Generic configurations of D4-branes stretched between and intersecting NS5-branes and D6-branes in $\mathbf{R}^{9,1}$ \cite{Witten:1997sc} preserve  4d $\mathcal N=2$ supersymmetry (eight supercharges). The low-energy theories on these branes are 4d gauge theories represented by linear quivers with $n$ special unitary gauge groups, $G = \prod_{i=1}^n SU(N_i)$, with bi-fundamental hypermultiplets
between adjacent groups\footnote{Adding orientifolds gives straightforward generalizations to orthogonal and symplectic groups, that we will not discuss here. Putting the D4-branes on a circle leads to circular quivers, whose gravitational duals are rather different and will not be discussed here, though there are
some limits in which these theories are related to ours.}, and extra fundamental hypermultiplets. In general these theories have only eight supercharges, making it difficult to find their gravity duals, since there is no classification of gravity solutions with eight supercharges; in particular, this is the case for the configuration of D4-branes ending on NS5-branes or D6-branes. However, when all D4-branes stretch between NS5-branes and/or D6-branes, and for a particular choice of the number of fundamental hypermultiplets for each gauge group, these brane configurations flow at low energies to superconformal linear quiver theories. The supersymmetry algebra for a conformal configuration is enhanced in the low-energy field theory to $SU(2,2|2)$, with 16 supercharges. These SCFTs have $n$ complex exactly marginal deformations, corresponding to the gauge couplings and theta angles.

The near-horizon limit of these brane configurations is some solution of type IIA string theory that is dual, by the AdS/CFT correspondence \cite{Maldacena:1997re,Gubser:1998bc,Witten:1998qj}, to these low-energy SCFTs. In some cases this solution may have a gravity approximation as a solution to type IIA supergravity. However, the solution may have regions where it is strongly coupled, so we may need to consider solutions to eleven dimensional supergravity (M theory); any solution of type IIA supergravity may be lifted to a solution of 11d supergravity, but the opposite is not necessarily true.
Thus, to look for duals of such superconformal linear quivers
which have a good gravity approximation, we need to look for
solutions to type IIA or 11d supergravity that preserve the 4d ${\cal N}=2$ superconformal algebra $SU(2,2|2)$ with 16 supercharges, and, in particular, with $SO(2,4) \times SO(3) \times U(1)$ isometry. The general solution of 11d supergravity preserving these (super)symmetries was found in \cite{Lin:2004nb} (see also \cite{OColgain:2010ev}) in terms of a function that satisfies the 3d continuous Toda equation. Using this result, the gravity dual of a large class of 4d $\mathcal N=2$ SCFTs (which includes the superconformal linear quiver theories we consider, and the SCFTs of class $\mathcal S$ introduced in \cite{Gaiotto:2009we}) was identified and analyzed by Gaiotto
and Maldacena \cite{Gaiotto:2009gz}. However, finding these duals explicitly (in the gravity approximation) generally involves solving the complicated non-linear Toda equation. As discussed in \cite{Gaiotto:2009gz}, this equation simplifies when the solution can be reduced to type IIA supergravity, in which case the 3d Toda equation reduces to the (linear) axially symmetric Laplace equation in three dimensions.

Since we start from a type IIA brane configuration, we expect that (at least for some range of parameters) our theories will have a good type IIA supergravity approximation. So,
in the present paper we consider the question: for which 4d SCFTs arising as the low-energy theories on configurations of D4-branes stretched between and intersecting NS5-branes and D6-branes is there a type IIA dual that is weakly coupled and weakly curved (away from brane sources) \footnote{In type IIB string theory it was proven in \cite{Colgain:2011hb} that no such weakly curved solutions, with the $SU(2)_R$ symmetry realized geometrically, exist.}? A special case of such type
IIA solutions was recently analyzed by Reid-Edwards
and Stefanski \cite{ReidEdwards:2010qs}. We show that these solutions describe (in some cases) smooth backgrounds with D6-branes and NS5-branes, and we show that they give a good approximation to the dual of the infinite coupling limit of linear superconformal quivers that have a large number of gauge groups, with large ranks for all groups. We then show how to go away from the infinite coupling limit. We find that
smooth solutions (away from branes) exist if we take finite couplings for a finite number
of gauge groups in the large quiver. The general solution of this type is given by solving an axially symmetric Laplace equation in three dimensions, with specific boundary conditions that encode the ranks of the gauge groups and the finite gauge couplings. We were not able to find explicit solutions to this equation in the general case, but we discuss some of their general properties, and we hope that (since a linear differential equation is involved) the solutions may be explicitly found in the future. This would enable explicit computations to be performed in these 4d ${\cal N}=2$ SCFTs at strong coupling. It would
also be interesting to generalize our constructions to cases with orientifolds, and to similar theories in other dimensions (in particular 2d and 5d SCFTs).

Note that one of the quiver theories we discuss is the $SU(N_c)$ gauge theory with $N_f=2N_c$, but for this case the duals we discuss are highly curved and cannot be trusted. The dual theory in this specific case was recently discussed in \cite{Gadde:2009dj,Gadde:2010zi} (see also \cite{Passerini:2011fe,Bourgine:2011ie,Fraser:2011qa}), where it was argued that it probably does not have a good gravitational approximation. However, we see that this theory can be viewed as a limiting case of theories (long
linear quivers with many gauge groups) that do have good gravitational dual descriptions. Hopefully, this will be useful for finding a dual string theory description for this theory (finding such a dual is challenging since when the 't Hooft coupling $g_{YM}^2 N_c$ is large, the open string coupling of the dual $g_s N_f$ is also large).

The paper is organized as follows. In section 2 we review the 4d $\mathcal N=2$ SCFTs that are realized as the low-energy limit of D4-branes ending on and intersecting NS5-branes and D6-branes. We provide the necessary conditions on the matter content for a theory to be conformal, and their geometrical interpretation in terms of the brane picture. In section 3 we review the general solution to 11d supergravity with $SU(2,2|2)$ symmetry, and how it relates to the supergravity duals to these 4d SCFTs. The solutions are parameterized by a function that satisfies the 3d Toda equation. We then review the reduction ansatz to type IIA for solutions with $U(1)$ isometry, in which case the Toda equation reduces to the axially symmetric Laplace equation in three dimensions.  In section 4 we review the subset of solutions of type IIA supergravity found explicitly in \cite{ReidEdwards:2010qs}. We discuss the field theory duals of these solutions, and analyze their range of validity. In section 5 we generalize to the case of finite
gauge couplings. We
construct the general solution of type IIA supergravity dual to linear quiver SCFTs in terms of a function that satisfies an axially symmetric Laplace equation in three dimensions with specific boundary conditions, and provide the necessary conditions for its validity.

\section{Review of 4d $\mathcal N=2$ SCFTs from brane configurations} \label{sec:field.theory}

In this section we review the 4d $\mathcal N=2$ SCFTs that are realized as the low-energy limit of D4-branes ending on and intersecting NS5-branes and D6-branes \cite{Witten:1997sc}. The low-energy theories on these branes are represented by linear quivers with special unitary gauge groups. We describe the necessary conditions on the matter content for a theory to be conformal, and their geometrical interpretation in terms of the brane picture. We end by summarizing the general expectations for the form of the holographic duals for these superconformal theories.

\subsection{4d $\mathcal N=2$ brane configurations in type IIA}

\begin{figure}[h]
  \centering
  \includegraphics[scale=.7]{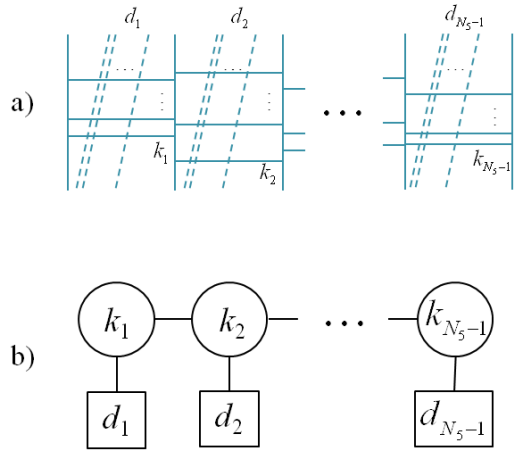}
      \caption{a) Canonical description of a general brane configuration. Horizontal lines denote D4-branes, vertical lines denote NS5-branes and diagonal broken lines denote D6-branes. b) Linear quiver describing the gauge theory corresponding to the brane configuration in a). Circular nodes denote special unitary gauge groups $SU(k_n)$. Horizontal lines between consecutive circular nodes denote bifundamental hypermultiplets. Square nodes denote a unitary global symmetry group $U(d_n)$. A line connecting a square and a circular node denotes $d_n$ hypermultiplets in the fundamental representation of $SU(k_n)$.}
      \label{fig:brane.conf}
\end{figure}

We consider configurations of D4-branes, NS5-branes and D6-branes in type IIA superstring theory on $\bf R^{9,1}$, preserving eight supercharges. The corresponding world-volumes span the coordinates $(x^0,x^1,x^2,x^3,x^6)$, $(x^0,x^1,x^2,x^3,x^4,x^5)$ and $(x^0,x^1,x^2,x^3,x^7,x^8,x^9)$, respectively,
with the D4-branes having a finite extent in the $x^6$ coordinate.
For the conformal theories that we are interested in, all D4-branes and NS5-branes are at $x^7=x^8=x^9=0$, and all D4-branes and D6-branes are at $x^4=x^5=0$.
%
%
By moving D6-branes in the $x^6$ direction (without changing the low-energy physics), we can bring any such supersymmetric brane configuration which describes a non-trivial supersymmetric quantum field theory (QFT) to a ``canonical form''. In this form we have
$N_5$ NS5-branes (with $N_5 \geq 2$ whenever we have a non-trivial QFT), with $k_n$ ($n=1,...,N_5-1$) D4-branes stretched between the $n^{th}$ and $(n+1)^{th}$ NS5-branes, and intersecting a stack of $d_n$ D6-branes. The total number of D6-branes is $\sum_{n=1}^{N_5-1} d_n =N_{D6}$; see figure \ref{fig:brane.conf} a).

By moving the NS5-branes and the D6-branes around in the $x^6$ direction, we can find different brane descriptions of the same low-energy physics. As explained in \cite{Ganor,Hanany:1996ie} for a similar configuration in type IIB string theory, when a D6-brane and a NS5-brane move past each other in the $x^6$ direction, a D4-brane stretched between them is created or annihilated. Brane configurations leading to inequivalent low-energy physics are characterized by a set of topological invariants for the NS5-branes and the D6-branes called the linking numbers \cite{Hanany:1996ie}. These invariants are defined up to constant shifts. There are two natural definitions for the linking numbers $K_i$ and $L_n$ of the $i^{th}$ NS5-brane, and of the $d_n$ D6-branes in the $n^{th}$ stack of D6-branes. Denoting by $D4^R_{n,i}$ the
number of D4-branes stretched to the right of the $i^{th}$ NS5-brane or $n^{th}$ stack of D6-branes, by
$D4^L_{n,i}$ the
number of D4-branes stretched to the left of the $i^{th}$ NS5-brane or $n^{th}$ stack of D6-branes, by
$D6^{R,L}_i$ the number of D6-branes to the right (left) of the $i^{th}$ NS5-brane, and by
$NS5^{R,L}_n$ the number of NS5-branes to the right (left) of the $n^{th}$ stack of D6-branes, these definitions are given by :
\be
\begin{array}{lll}
1) & {\tilde K}_i = D4^R_i - D4^L_i + D6^L_i , &  {\tilde L}_n = D4^R_n - D4^L_n - NS5^R_n  ;\\
&& \\
2) & K_i = D4^R_i - D4^L_i - D6^R_i , &  L_n = D4^R_n - D4^L_n + NS5^L_n.\\
\end{array}
\e
These two definitions are convenient because they both obey that the sum of all linking numbers vanishes :
\be \label{eq:linking.sum.zero}
\sum_{i=1}^{N_5}K_i+\sum_{n=1}^{N_5-1}d_nL_n = \sum_{i=1}^{N_5}{\tilde K}_i+\sum_{n=1}^{N_5-1}d_n{\tilde L}_n = 0.
\e

Since the D4-branes have finite extension in the $x^6$ direction, the low-energy physics on the D4-branes is four-dimensional. The 4d low-energy gauge theory can be read off from the brane configuration. For $k_n$ coincident D4-branes, the D4-D4 open strings for a given $n$ are described by an $\mathcal N=2$ super Yang-Mills (SYM) theory with gauge group $U(k_n)$. The $U(1)$ factor decouples at low energies (from the brane point of view, it is non-dynamical due to its coupling to the NS5-branes \cite{Witten:1997sc}). The total gauge group of the configuration is thus $\prod_{n=1}^{N_5-1} SU(k_n)$.
The D4-D4 strings between the $k_{n-1}$ and $k_n$ D4-branes ending on the $n^{th}$ NS5-brane give rise to a massless hypermultiplet transforming in the bifundamental representation $\bf{(k_{n-1},\bar k_n)}$.
In addition, the D4-D6 strings stretched between the $k_n$ D4-branes and the $d_n$ D6-branes contribute $d_n$ massless hypermultiplets in the fundamental representation $\bf{k_n}$.
 The field theory corresponding to the given brane configuration is thus the 4d $\mathcal N=2$ gauge theory represented by the linear quiver diagram in figure \ref{fig:brane.conf} b).

The one-loop beta function coefficient of the gauge node $SU(k_n)$ is
\be
b_{0,n}=-2k_n+k_{n-1}+k_{n+1}+d_n.
\e
%
A $\mathcal N=2$ gauge theory is conformal if and only if all these beta function coefficients vanish,
\be \label{eq:conformal.condition}
d_n= 2k_n-k_{n-1}-k_{n+1}, \qquad \forall~n
\e
(with $k_0 = k_{N_5} = 0$).
In particular, configurations of this type without D6-branes can not be conformal (\eqref{eq:conformal.condition} implies that $N_{D6} = k_1 + k_{N_5-1}$). The supersymmetry algebra for a conformal configuration is enhanced in the low-energy field theory to $SU(2,2|2)$, with 16 supercharges. We will be interested in these superconformal brane configurations in the present paper. The condition \eqref{eq:conformal.condition} implies that the linking numbers of the NS5-branes are all the same, and given by (using the second definition, as we will from here on)
\be\label{nsfivelink}
K_i=-k_{N_5-1}, \qquad  i = 1,2,\cdots,N_5.
\e

In generic brane configurations of the type described above, the D4-branes ending on the NS5-branes
causes them to bend logarithmically ($x^6$ depends logarithmically on $|x^4+ix^5|$) \cite{Witten:1997sc}.
More precisely, the coefficient of the logarithmic bending is governed by the linking number of
each NS5-brane. Such a bending breaks the low-energy conformal symmetry so it cannot appear
in superconformal brane configurations. Indeed, for such configurations in which the linking numbers of
all NS5-branes are equal, there is a coordinate system in which there is no bending, and each
NS5-brane sits at a fixed value $x_i^6$ ($i=1,\cdots,N_5$). The D6-branes also bend by an
amount proportional to their linking number, but this bending goes as a power law (since the
D4-branes are co-dimension three objects inside the D6-branes) so it is consistent with the
low-energy conformal invariance.

The classical gauge coupling $g_n$ of the gauge group $SU(k_n)$ is obtained from the 5d coupling of the world-volume theory of the $k_n$ D4-branes by a Kaluza-Klein reduction over the interval $[x_n^6,x_{n+1}^6]$. The gauge couplings for a superconformal linear quiver are given by
\be\label{gauge_couplings}
\frac{1}{g_n^2} = \frac{x_{n+1}^6-x_n^6}{g_s l_s}.
\e
The low-energy limit giving our SCFTs is described by taking $l_s$ to zero while keeping fixed $g_n^2$
given by \eqref{gauge_couplings}. In this limit all NS5-branes are very close to each other in string units. We can further take a
strong coupling limit $g_n \rightarrow \infty$ for the $SU(k_n)$ gauge group, by bringing together the two NS5-branes, $x_{n+1}^6 \rightarrow x_n^6$. We will see that such strong coupling limits will play a special role in the holographic duals that we will find. Note that a given configuration can have different strong coupling limits, in which $m<N_5$ NS5-branes are kept at a finite distance, i.e. the gauge couplings of $m-1$ gauge groups are kept finite, while the remaining gauge couplings are sent to infinity.

The theta angle $\theta_n$ of the gauge node $SU(k_n)$ is not visible geometrically in the type IIA brane configuration, but it becomes visible when it is lifted to M-theory \cite{Witten:1997sc}. Recall that type IIA superstring theory on $\bf R^{9,1}$ is equivalent to M-theory on ${\bf R^{9,1}} \times S^1$, and NS5-branes lift to M5-branes localized on the $S^1$. If we denote the $S^1$ coordinate by $x^{10}$ (with periodicity $2\pi$), the theta angle of the $n^{th}$ gauge group is the difference between the locations of the $(n+1)^{th}$ and $n^{th}$ NS5-branes in the M-theory circle, $\theta_n = x^{10}_{n+1} -x_{n}^{10}$.

To summarize, a linear ${\cal N}=2$ superconformal quiver theory of the type we consider in this
paper is parameterized by the number of NS5-branes $N_5$ (equal to the number of gauge group factors plus one), the $(N_5-1)$ complexified coupling constants (including the gauge couplings $g_n$ and the theta angles $\theta_n$), and the integer parameters $d_n$ and $L_n$ determining the number of D6-branes in each D6-brane stack and their linking numbers. From these numbers one can compute the $k_n$, using \eqref{eq:conformal.condition}, and thus identify the gauge groups.
The $L_n$'s are constrained by \eqref{eq:linking.sum.zero} (noting that $K_i$ are determined using \eqref{nsfivelink} in terms of $k_n$).

\subsection{Examples of 4d $\mathcal N=2$ SCFTs} \label{subsec:Examples}

We proceed by giving examples of two classes of superconformal linear quiver gauge theories that
have interesting large $N$ limits that we will examine later in this paper. Generally a large $N$
limit involves a large number of D4-branes, but the numbers of other branes can be kept finite or
grow with $N$, depending on the case.

\begin{figure}[h]
  \centering
  \includegraphics[scale=.7]{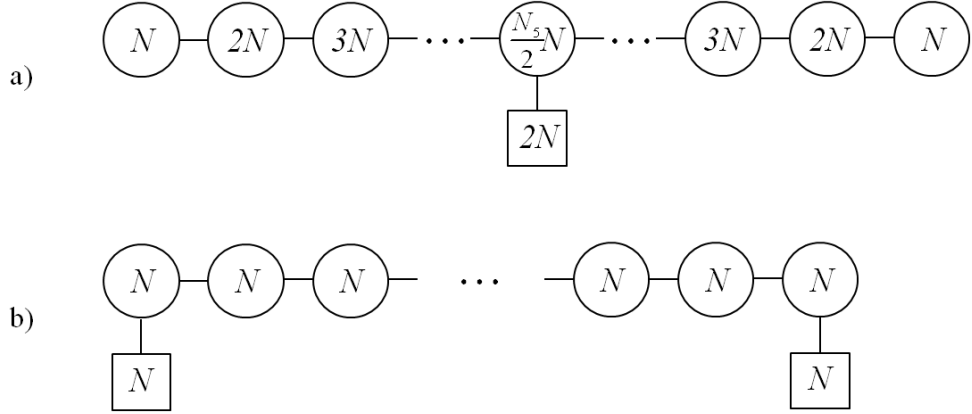}
      \caption{a) The quiver diagram for our first example, with varying ranks. b) The quiver diagram for our second example, with equal ranks.}
      \label{fig:conf.quiver}
\end{figure}

\begin{enumerate}

\item Example with different gauge group ranks and one stack of D6-branes :

Consider the following theory (for even values of $N_5$) :
%
\be
\begin{split}
& k_n = \left\{ \begin{array}{ll}
                nN, & n=1,\cdots,N_5/2 \\
                (N_5-n) N,  &  n=N_5/2+1,...,N_5-1, \\
                \end{array} \right.\\
& d_n = \left\{ \begin{array}{ll}
                2N,& n = N_5/2 \\
                0, &  n \ne N_5/2.\\
                \end{array} \right. \\
\end{split}
\e
In this theory we can obtain large rank gauge groups either by taking $N$ to be large,
in which case the number of D6-branes becomes large but the number of NS5-branes does not,
or by taking $N_5$ to be large, in which case the number of NS5-branes becomes large but the
number of D6-branes does not.
 The associated quiver is depicted in figure \ref{fig:conf.quiver} a). The linking numbers are
\be
\begin{split}
& K_i = -N \qquad i=1,\cdots,N_5,\\
& L_{N_5/2} = N_5/2.\\
\end{split}
\e
For the special case of $N_5=2$ this theory is simply the $SU(N)$ SQCD theory with $2N$ flavors.

\item Example with equal gauge group ranks and two stacks of D6-branes :

This example is defined by
\be
\begin{split}
& k_n = N, \qquad n=1,...,N_5-1,\\
& d_n = \left\{ \begin{array}{ll}
                 N, & n=1,N_5-1 \\
                 0, & n=2,...,N_5-2.\\
                \end{array} \right.\\
\end{split}
\e
Here there is only one large $N$ limit one can take, by taking the numbers of D4-branes and
D6-branes to infinity while keeping the number of NS5-branes fixed.
The associated quiver is depicted in figure \ref{fig:conf.quiver} b). The linking numbers are
\be
\begin{split}
& K_i = -N,  \qquad i= 1,...,N_5,\\
& L_n = \left\{ \begin{array}{ll}
                1, & n=1 \\
                (N_5-1), &  n = N_5-1.\\
                \end{array}\right.\\
\end{split}
\e
For $N_5=2$ this theory is again equivalent to the $SU(N)$ SQCD theory with $2N$ flavors,
but for higher values of $N_5$ the two examples yield different theories.

\end{enumerate}

Of course, in all examples the linking numbers satisfy \eqref{eq:linking.sum.zero}.

\subsection{Expectations for the dual string theory backgrounds}

In this paper we would like to find the string theory backgrounds dual to the superconformal theories
described above. On general grounds \cite{Maldacena:1997re,Gubser:1998bc,Witten:1998qj}, we expect these to be given by the near-horizon limit of the brane configurations of figure \ref{fig:brane.conf}\ a). The near-horizon limit replaces the D4-branes by their
back-reacted background, while leaving the NS5-branes and D6-branes as branes in the background, which may or may not have large back-reactions. In particular, the gauge fields on the D6-branes in the holographic dual backgrounds provide the flavor symmetries acting on the fundamental hypermultiplets for our superconformal theories, so we should keep these gauge fields in our description.

The most naive expectation would be to get a background of type IIA string theory.
%
The bosonic symmetry $SO(2,4) \times SU(2)_R\times U(1)_R$ of our theories (of which an $SO(1,3)\times SU(2)_R\times U(1)_R$ subgroup is realized geometrically in the brane construction) requires the dual metric to be a product $AdS_5 \times S^2 \times S^1 \times \Sigma'$, with the other spaces warped over the Riemann surface $\Sigma'$. The NS5-branes and D6-branes are expected to wrap $AdS_5 \times S^1$ and the $AdS_5 \times S^2$ subspaces of this, respectively. This is completely analogous to the recent constructions of type IIB supergravity solutions dual to D3-branes ending on 5-branes \cite{Aharony:2011yc} and D3-branes stretched between 5-branes \cite{Assel:2011xz} (these constructions are special cases of the general solutions of \cite{D'Hoker:2007xy,D'Hoker:2007xz}).

However, it is not clear if this expectation is correct. First, it is far from obvious that the string
coupling in the holographic description is weak, and if not, we would need to lift the backgrounds above to M theory, and replace $\Sigma'$ by some three-dimensional manifold. If we were taking the 't Hooft limit of large
number of colors with $g_{YM}^2 N$ fixed and the numbers of flavors fixed, we would expect to get a weakly coupled string background (with $g_s \sim 1/N$). However, in the most naive large $N$ limit, where we keep the number of NS5-branes fixed, the number of D6-branes grows with the number of D4-branes, so this is a Veneziano-type limit in which the number of flavors scales with the number of colors, and it is not clear if it should really have a weakly coupled string description or not.  In particular, for our second example there is no apriori reason to expect the string coupling in the dual description to be weak. In our first example, if we take large $N$ with fixed $N_5$, we have a large number of flavors at large $N$, but for large $N_5$ it is much smaller than the number of colors, so we may expect a weakly coupled string dual in this case (which is close to the usual 't Hooft limit). If, on the other hand, we take in this example large $N_5$ instead of large $N$, then the number of flavors is small, but the number of gauge groups is very large, so it is again not clear if a weakly coupled string description should exist.

Similarly, it is not clear apriori when the dual string backgrounds should be weakly curved. In the standard 't Hooft limit, this often happens when the 't Hooft couplings $g_{YM}^2 N$ are large. This is the case in particular for the ${\mathcal N}=2$ theories coming from circular quivers. But for most of our theories the number of flavors is large, so it is not obvious when the curvature should be small. A similar concern is that in order to be able to trust string perturbation theory, we not only need $g_s$ to be small in the dual description, but also the couplings on the D6-branes and the NS5-branes. The former go as $g_s d_n$, which is one of the reasons why we need the number of flavors to be much smaller than the number of colors.

Unfortunately, when the dual background is not weakly curved we do not have any good tools to analyze it. In this paper we will look for solutions that are weakly curved, and check which gauge theories (for which parameters) are well-described by them. For simplicity we will also look only for solutions in type IIA string theory, and not in M theory. This is expected to include any 't Hooft-like limits of our gauge theories, and it may include also additional theories. Other limits may have good descriptions in M theory, that are beyond the scope of this paper. Since we are looking for weakly curved backgrounds in type IIA string theory, we will be analyzing solutions to type IIA supergravity. For each solution we will need to check if the solution is weakly coupled and curved, and if the theories on the branes are also weakly coupled, for our approximations to be self-consistent. We will see that indeed there is some class of theories, and some range of couplings, for which such type IIA supergravity backgrounds (with branes) will be good approximations to the holographic duals of our SCFTs.


\section{Review of M-theory solutions with $SU(2,2|2)$ symmetry}

\subsection{11d supergravity with $SO(2,4) \times SO(3) \times U(1)$ symmetry}

The general solutions of 11d supergravity preserving the 4d ${\cal N}=2$ superconformal algebra $SU(2,2|2)$ with 16 supercharges, and, in particular, with $SO(2,4) \times SO(3) \times U(1)$ isometry, were found in \cite{Lin:2004nb}. Any solution of type IIA
supergravity can be lifted to an M theory solution with no dependence on the M theory circle, so,
as we will discuss in more detail below, the type IIA solutions we are looking for should be special
cases of these solutions. As discussed above, such solutions contain a three-dimensional space whose
form is not determined by the isometry, and we label it by $(x_1, x_2, y)$.
A solution with no sources is then
described \cite{Lin:2004nb} by a single function $D(x_1,x_2,y)$ that obeys the continuous Toda equation
\be \label{eq:Toda}
\partial_{x_1}^2 D+ \partial_{x_2}^2 D + \partial_{y}^2 e^D =0.
\e
The 11d supergravity fields take the form
\be
\begin{split} \label{eq:11d.fields}
ds_{11}^2 = & \kappa^{2/3} e^{2 \lambda} \left[ 4 ds_{AdS_5}^2 + y^2 e^{-6 \lambda} ds_{S^2}^2 + \frac{4}{1-y \partial_yD}\left(d\chi + v_i dx^i\right)^2 - \frac{\partial_y D}{y} \left(dy^2 + e^D \delta_{ij}dx^i dx^j\right) \right], \\
G_{(4)}= & dC_{(3)} = \kappa F_{(2)} \wedge d \Omega_2,
\end{split}
\e
where
\be
\begin{split}
F_{(2)} = & 2 \left[(d\chi+v_idx^i) \wedge d (y^3 e^{-6 \lambda}) + y (1-y^2 e^{-6 \lambda}) dv_i \wedge dx^i - \frac{1}{2} \partial_y e^D dx^1 \wedge dx^2 \right], \\
v_i = & \frac{1}{2} \varepsilon_{ij}\partial_j D, \ (i=1,2)\qquad \qquad e^{-6 \lambda} = - \frac{\partial_yD}{y (1-y \partial_y D)}.
\end{split}
\e
Here, $d\Omega_2$ is the volume form on the two-sphere.
The overall normalization constant $\kappa$ determines the normalization of the 4-form flux, and can be rescaled by rescaling $y$ and $e^D$.

These solutions were analyzed in \cite{Gaiotto:2009gz}, where the authors identified the necessary boundary conditions for the Toda equation to obtain supergravity duals to the large class of 4d $\mathcal N=2$ SCFTs constructed in \cite{Gaiotto:2009we} (which include the theories we discussed in the previous section). In the rest of this section, we summarize these conditions for the case of supergravity solutions dual to the subset of $\mathcal N=2$ SCFTs described in section \ref{sec:field.theory}.

\subsection{Boundary conditions for the Toda equation}

Regularity of the metric at the point $y=0$, where $S^2$ shrinks to zero size, requires
\be \label{eq:b.c.orig}
\partial_y D |_{y=0} =0, \qquad \qquad e^D|_{y=0} = \textrm{finite},
\e
such that the last term in the metric remains finite ($\partial_y D / y$ has a smooth limit as $y \to 0$).

As discussed above we expect our solutions to describe theories on D4-branes, and to contain NS5-branes, which means that when lifted to M-theory we expect them to include non-trivial 4-form flux.
In order to have a non-trivial 4-cycle that can support 4-form flux, we look for solutions where $S^1_{\chi}$ shrinks to zero size at a point $y=y_c$, such that a 4-cycle $M_4$ is given by the product $S^2 \times S^1_{\chi} \times I_y$  warped over the interval $I_y = \{y| 0 \le y \le y_c \}$. This requirement is equivalent to
\be
\partial_y D  \xrightarrow{y \rightarrow y_c} \infty,
\e
which implies
\be
e^{-6\lambda} |_{y=y_c} = \frac{1}{y_c^2}.
\e
Regularity of the metric at $y=y_c$ then requires
%
\be
e^D |_{y=y_c} \sim (y - y_c),
\e
and the 4-form flux on $M_4$ is then
\be
\begin{split}
\int_{M_4} G_{(4)} = & \kappa \int_{S^2} d\Omega_2 \int_{S^1_{\chi} \times I_y} F_{(2)}|_{x^i = \textrm{const}} \\
= & \kappa \int_{S^2} d\Omega_2 \int_{S^1_{\chi}} d\Omega_1 \int_0^{y_c} dy ~ 2 \partial_y (y^3 e^{-6 \lambda}) \\
= & \kappa (4 \pi)^2 y_c.
\end{split}
\e
Choosing $\kappa = \frac{\pi}{2} l_p^3$
implies that $y_c$ is quantized to be an integer.

We want to allow also sources to the supergravity equations,
corresponding to NS5-branes in type IIA and thus to M5-branes in M theory.
The presence of a localized source for the 3-form potential with $N_5$ units of 4-form flux requires that the Toda equation be supplemented with an extra boundary condition in the form of a singular source located at $x^i=x^i_0$ and extended along $y$ as follows \cite{Gaiotto:2009gz}
\be \label{toda_source}
\partial_{x_1}^2 D+ \partial_{x_2}^2 D + \partial_y^2 e^D = -2\pi \delta^{(2)}(x-x_0)\theta(2N_5-y).
\e
Although solutions to \eqref{toda_source} are singular, the full ten dimensional metric is non-singular.

\subsection{Solutions with an extra $U(1)$ symmetry}

Finding solutions to the non-linear Toda equation \eqref{eq:Toda} is a difficult task.
However, as mentioned above, we are mostly interested in type IIA solutions which would describe
the standard 't Hooft limit (if it exists) of the gauge theories we are interested in. Formally
we can reduce a solution to type IIA if it is independent of some compact circle direction, so, following \cite{Gaiotto:2009gz,ReidEdwards:2010qs}, we
will look for solutions that have such an isometry. Note that since we have sources, this means that we will have to smear our sources along the isometry directions, and our solutions will not correspond directly to physical theories in which these sources are localized. Specifically, since the positions of the branes in the M theory direction determine the theta angles, our solutions will involve a smearing over all values of these parameters. Apriori one would not expect such a smearing to give sensible results, but we will argue below that in some range of parameters it does, and that the smeared solutions we will find are
good approximations to the exact solutions, in which the sources are not smeared (and there is no isometry).

So, we consider polar coordinates $(r,\beta)$ in the $x_1x_2$-plane, and look for solutions with a $U(1)_{\beta}$ isometry $D(x_1,x_2,y)=D(r,y)$. It turns out \cite{Ward:1990qt} that in this case the problem simplifies by performing the following implicit change of coordinates from $(r,y)$ to $(\rho,\eta)$,
\be \label{eq:coord.transf}
r^2 e^D = \rho^2, \qquad y = \rho \partial_{\rho} V \equiv \dot V, \qquad \ln r = \partial_{\eta} V \equiv V',
\e
which replaces $D(r,y)$ by $V(\rho,\eta)$.
In terms of $V$, the Toda equation becomes a two-dimensional partial differential equation (PDE) that is equivalent to the axially symmetric Laplace equation in three dimensions
\be \label{eq:Laplace}
\frac{1}{\rho} \partial_{\rho}(\rho \partial_{\rho}V)+\partial_{\eta}^2V=0, \qquad \textrm{or} \qquad \ddot V + \rho^2 V'' = 0.
\e

In the new variables, the 11d supergravity fields \eqref{eq:11d.fields} take the form
\be
\begin{split} \label{eq:11d.solution}
ds_{11}^2 = & \kappa^{2/3} \left(\frac{\dot V \tilde \Delta}{2V''} \right)^{1/3} \left[ 4 ds_{AdS_5}^2 + \frac{2V'' \dot V}{\tilde \Delta}ds_{S^2}^2 + \frac{2(2\dot V - \ddot V)}{\dot V \tilde \Delta} \left(d\beta + \frac{2\dot V \dot V'}{2\dot V - \ddot V} d\chi \right)^2 + \right. \\
& \left. ~~~~~~~~~~~~~~~~~~~~ + \frac{4V''}{2 \dot V - \ddot V} \rho^2 d\chi^2 + \frac{2V''}{\dot V} (d\rho^2 + d\eta^2) \right],\\
C_{(3)} = &  2 \kappa \left[ -2 \frac{\dot V^2 V''}{\tilde \Delta} d\chi + \left( \frac{\dot V \dot V'}{\tilde \Delta} - \eta \right) d\beta \right] \wedge d\Omega_2,
\end{split}
\e
where
\be
\tilde \Delta \equiv (2 \dot V - \ddot V)V'' + (\dot V')^2.
\e

\subsection{Boundary conditions for the Laplace equation} \label{sec:b.c.V}

In the new coordinates, the $y=0$ plane maps to any surface on which ${\dot V}$ vanishes,
while $y=y_c$ maps to $\rho=0$. The $\rho$ coordinate goes from zero to infinity, diverging
whenever we have a 5-brane source \eqref{toda_source}. On the other hand, the $\eta$ coordinate
turns out to have finite extent (which we take to be from $\eta=0$ to $\eta=\eta_c$) in interesting
solutions. Thus, the remaining two-dimensional space $\Sigma'$ is a half-strip $0 < \rho < \infty, 0 < \eta < \eta_c$. As discussed in \cite{Gaiotto:2009gz}, one can have orbifold singularities in M theory at $r=0$ or at $r=\infty$, which will map to D6-branes in type IIA; in the new coordinates these sit at $\rho=0$ and some value of $\eta$.

The boundary conditions for the two-dimensional PDE \eqref{eq:Laplace} in terms of $\dot V(\rho,\eta)$ become
\be \label{eq:b.c.}
\begin{split}
\dot V |_{\eta = 0, \eta_c} =0, \qquad \dot V|_{\rho = 0} = \lambda(\eta),
\end{split}
\e
with $\lambda(\eta)$ related to the D6-branes.
If we think of \eqref{eq:Laplace} as an axially symmetric electrostatics problem in three dimensions, then
$\lambda(\eta)$ can be interpreted as a {\it line charge density} sourcing the {\it electrostatic} potential $V(\rho,\eta)$, while the first boundary condition can be interpreted as two {\it infinite conducting disks} located at $\eta =0, \eta_c$. In this language, the potential $V$ satisfies the axially symmetric Poisson equation in three dimensions
\be \label{eq:Poisson}
\frac{1}{\rho} \partial_{\rho}(\rho \partial_{\rho}V)+\partial_{\eta}^2V = \lambda(\eta) \frac{1}{\pi |\rho|}\delta(\rho),
\e
subject to the boundary conditions $\dot V |_{\eta = 0, \eta_c} =0$.

\begin{figure}[h]
  \centering
  \includegraphics[scale=.7]{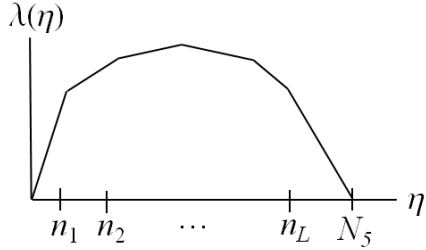}
      \caption{Typical consistent profile for the line charge density $\lambda(\eta)$.}
      \label{fig:profile}
\end{figure}

Quantization of fluxes wrapping non-trivial cycles imposes strong constraints on the allowed form of $\lambda(\eta)$ \cite{Gaiotto:2009gz}. They can be summarized as follows:

\begin{itemize}

\item The line charge density $\lambda(\eta)$ must be piece-wise linear and continuous, composed of segments of the form $\lambda(\eta) = a_i \eta + q_i$ with $a_i \in \mathbb{Z}$.

\item The change of gradient at a kink (where two adjacent line segments meet) must be an integer and the gradient must decrease with successive line segments, i.e. $a_i-a_{i-1} \in \mathbb{Z}^-$.

\item $\lambda(0)= \lambda(\eta_c)=0$.

\item The positions of the kinks  in the $\eta$-axis must be at integer values $\eta=n_i \in \mathbb{Z}^+$.

\end{itemize}
This implies in particular that $\eta_c=N_5 \in \mathbb{Z}^+$ and $\eta \in [0,N_5]$; it is easy to show that $N_5$ is the 4-form flux emanating from the sources at $\rho=\infty$. We will analyze in more detail the sources at $\rho=\infty$ below. A typical $\lambda$-profile is depicted in figure \ref{fig:profile}.
When $|a_i-a_{i-1}| >1$ the M theory geometry contains an $A_{|a_i-a_{i-1}|-1}$ singularity transverse to $AdS_5 \times S^2$,
which will reduce to D6-branes in type IIA. As we will see in section \ref{sec:1NS5}, this fact fits nicely with the brane interpretation of the solutions.


\subsection{Reduction to type IIA}

Having the $U(1)_{\beta}$ isometry, the solution \eqref{eq:11d.solution} can be reduced to type IIA supergravity by dimensionally reducing along the circle $S^1_{\beta}$. Using the standard reduction ansatz
\be
\begin{split}
ds_{11}^2 = & e^{- \frac{2}{3}\phi} g_{\mu \nu} dx^{\mu}dx^{\nu} + e^{\frac{4}{3} \phi} (d \beta + A_{\mu} dx^{\mu})^2, \\
C_{(3)} = & A_{(3)} + B_{(2)} \wedge d\beta ,
\end{split}
\e
one finds that the type IIA metric is the product $AdS_5 \times S^2 \times S^1 \times \Sigma'$ warped over $\Sigma'$, where $\Sigma'$ is the half-strip parameterized by the coordinates $(\rho,\eta)$. The explicit form of the metric in the string frame is
\be \label{eq:10d.metric}
\begin{split}
ds_{10}^2 = & \kappa^{2/3}\left( \frac{2 \dot V - \ddot V}{V''} \right)^{\frac{1}{2}} \left[ 4 ds_{AdS_5}^2 + \frac{2 V'' \dot V}{\tilde \Delta} ds_{S^2}^2 + \frac{2 V''}{\dot V} (d\rho^2 + d\eta^2) + \frac{4 V''}{2\dot V - \ddot V} \rho^2 d\chi^2 \right].\\
\end{split}
\e
The remaining NS-NS fields are
\be
e^{4 \phi} = \frac{4(2 \dot V - \ddot V)^3}{V'' \dot V^2 \tilde \Delta^2}, \qquad B_{(2)}= 2 \kappa^{2/3} \left(\frac{\dot V \dot V'}{\tilde \Delta} - \eta \right) d \Omega_2,\\
\e
while the RR potentials are
\be
A_{(1)} = \kappa^{1/3}\frac{2\dot V \dot V'}{2 \dot V - \ddot V} d \chi, \qquad A_{(3)} = - \kappa \frac{4 \dot V^2 V''}{\tilde \Delta} d \chi \wedge d\Omega_2.\\
\e
Locally this is a good solution whenever the metric is positive, and one can show that this is
true whenever
%
\be \label{eq:metric+condition}
\frac{\dot V}{V''} >0.
\e
We will analyze the singularities in these solutions, and their range of validity, below.

Following the conventions of \cite{Grana:2005jc}, quantization of various fluxes requires $\kappa = (\alpha')^{3/2}$. In the rest of the paper we will use units in which $\alpha'=1$.


\section{Solutions with one stack of NS5-branes} \label{sec:1NS5}

\subsection{General solution}

 A large class of solutions satisfying the boundary conditions of section \ref{sec:b.c.V} was constructed in \cite{ReidEdwards:2010qs}. We will see below that these are not the most general solutions with these boundary conditions, but we start by analyzing in detail the solutions that they found. A solution is characterized by a line charge density $\lambda(\eta)$ satisfying the conditions reviewed above,
%
\be \label{eq:gen.lambda}
\lambda(\eta) =  \left\{ \begin{array}{ccrcl}
                          a_0\eta & \textrm{if} & 0 \leq & \eta & \leq n_1 \\
                          a_1\eta+q_1    & \textrm{if} & n_1 \leq & \eta &  \le n_2 \\
                          \vdots & & & \\
                          a_L \eta + q_L & \textrm{if} & n_L \leq & \eta &  \le N_5 ,\\
                          \end{array}
\right.
\e
with $q_i$ determined in terms of $a_i, n_i$ by continuity of $\lambda(\eta)$.
The $2L+1$ parameters $a_i, n_i$ are constrained by the requirement $\lambda(N_5)=0$, or
%
\be \label{eq:lambda.const}
a_0 n_1 + a_1 (n_2-n_1)+a_2(n_3-n_2)+ ... + a_L (N_5-n_L)=0.
\e
The line charge density can be written as a sum of simple building blocks
\be \label{simple_comb}
\lambda(\eta) = - \sum_{i=1}^{L+1} (a_{i}-a_{i-1}) {\tilde \lambda}(\eta;n_i),
\e
where
\be \label{simple_source}
{\tilde \lambda}(\eta;n_i)\equiv \frac{1}{2} \left( |\eta+n_i| - |\eta - n_i| \right)
                    =\left\{ \begin{array}{ccrcl}
                       \eta & \textrm{if} & 0 \leq & \eta & \leq n_i \\
                       n_i     & \textrm{if} & n_i \leq & \eta &  .\\
                       \end{array}
\right.
\e
In this expression $a_{L+1}=0$ and $n_{L+1}=N_5$. The solution may thus be written as a superposition
of the solutions with the simple sources \eqref{simple_source}.

The solution can be found by using the method of images with respect to the points $\eta=0,N_5$, which implies that the solution should be odd and periodic in $\eta$, with a half period given by the profile above, and extended along the whole real line. $\lambda(\eta)$ as written in \eqref{simple_comb} is odd, and the periodicity of $2N_5$ may be implemented by performing the replacement
\be
\lambda(\eta) \rightarrow \lambda_p (\eta) \equiv \sum_{m=-\infty}^{\infty} \lambda(\eta - 2mN_5).
\e
Due to the form of the boundary conditions, it is convenient to present the solution in terms of $\dot V$. The solution satisfying the boundary conditions \eqref{eq:b.c.} for the line charge density \eqref{eq:gen.lambda} can be written in two useful representations \cite{ReidEdwards:2010qs}
\be \label{eq:gen.sol}
\begin{split}
\dot V(\rho,\eta)= & \frac{1}{2} \sum_{m=-\infty}^{\infty} \sum_{i=1}^{L+1} (a_i - a_{i-1}) \left( \sqrt{ \rho^2 + (\eta - 2mN_5 - n_i)^2} - \sqrt{\rho^2 + (\eta - 2mN_5 + n_i)^2 } \right) \\
= &  \sum_{n = 1}^{\infty}\frac{\rho}{n} A_n K_1 \left( \omega_n \rho \right) \sin \left( \omega_n \eta \right). \\
\end{split}
\e
In the second equation $K_1$ denotes the modified Bessel function of the second kind of order one, and we defined
\be \label{eq:An}
A_n = \frac{2}{\pi}\sum_{i=1}^{L} (a_i - a_{i-1}) \sin \left( \omega_n n_i \right), \qquad \omega_n = \frac{n \pi}{N_5}.
\e
Note that the possible contribution with $n_{L+1} = N_5$ vanishes in the definition of $A_n$.


For consistency, the solution \eqref{eq:gen.sol} must satisfy the positivity condition \eqref{eq:metric+condition}. $V''$ satisfies the same differential equation \eqref{eq:Laplace} as $V$ does, while $\dot V$ satisfies a slightly different differential equation
\be
\partial_{\rho}^2 \dot V - \frac{1}{\rho}\partial_{\rho} \dot V + \partial_{\eta}^2 \dot V =0,
\e
 which is an elliptic PDE in two dimensions. {\it Hopf's maximum principle} \cite{Hopf:1927} for the solutions to an elliptic PDE says that extrema of any solution can only appear on the boundary.
The boundary conditions for $\dot V$ define the solution and are given by
\be \label{eq:b.c.Vdot}
\dot V|_{\rho=0}=\lambda(\eta)>0, \qquad \dot V|_{\eta=0,N_5}=0, \qquad \dot V \xrightarrow{\rho \rightarrow \infty} 0.
\e
Thus, $\dot V > 0$ in the interior. $V''$ satisfies a 2d equation that is equivalent to the axially symmetric Laplace equation in three dimensions, which is also elliptic, so the same principle holds there as well. From the 3d point of view the boundaries are at
$\rho\rightarrow \infty$ and at $\eta=0,N_5$, where $V''$ obeys
%
\be
V''|_{\eta=0,N_5}=0, \qquad V'' \xrightarrow{\rho \rightarrow \infty} 0.
\e
$V''$ also has sources (from the 3d point of view) in the interior, at $\rho=0$ and $\eta=n_i$, which are all positive (as we show explicitly below). Thus, we also have $V'' > 0$ everywhere, and the metric is indeed positive.

\subsection{Behavior near special points}

To understand the topology of the solution, we analyze the behavior of the supergravity fields near the zeros and singular points of the potential $V(\rho,\eta)$ and its various derivatives. We first analyze the metric and 3-form, and then tackle the problem of defining and computing a globally well-defined and conserved 4-form flux, which is subtle
due to the presence of a Chern-Simons type term in the type IIA supergravity action.
From the form of the metric \eqref{eq:10d.metric} and our boundary conditions, it is clear that the
2-sphere shrinks at $\eta=0,N_5$, and there is a non-trivial 3-cycle formed by this two-sphere together with the $\eta$ direction. We show that this 3-sphere surrounds NS5-branes sitting at $\rho=\infty$. We also show that at each point $\rho=0,\eta=n_i$ we have a collection of D6-branes surrounded by a non-trivial 2-cycle, and that all other points in the background are regular.

\subsubsection{Asymptotic region at $\rho \to \infty$}

 To study the behavior of the various derivatives of the potential $V(\eta,\rho)$ in the asymptotic region $\rho \rightarrow \infty$, we note that as $\rho \to \infty$
\be
K_1 \left( \omega_n \rho \right) \rightarrow \sqrt{\frac{\pi}{2 \omega_n \rho}} e^{-\omega_n \rho}.
\e
Thus, the leading term in the second line of \eqref{eq:gen.sol} is the one with $n=1$. Hence
\be
\begin{array}{ll}
\dot{V} \rightarrow \sqrt{\frac{\pi}{2 \omega_1}} A_1 \sin\left( \omega_1 \eta \right) \rho^{1/2} e^{-\omega_1 \rho}, &  V''  \rightarrow \sqrt{\frac{\pi \omega_1}{2}} A_1 \sin\left( \omega_1 \eta \right) \rho^{-1/2} e^{-\omega_1 \rho},\\
& \\
\dot{V}'  \rightarrow \sqrt{\frac{\pi \omega_1}{2}} A_1 \cos\left( \omega_1 \eta \right) \rho^{1/2} e^{-\omega_1 \rho}, & \tilde{\Delta} \rightarrow \frac{\pi}{2} \omega_1 A_1^2 \rho e^{-2 \omega_1 \rho}.\\
\end{array}
\e
The string frame metric then has the asymptotic form (for $\rho \to \infty$)
\be
ds_{10}^2 \rightarrow 4 \rho(ds^2_{AdS_5}+d\chi^2)+\frac{2 N_5}{\pi}\left[\sin^2\left(\frac{\pi \eta }{N_5}\right)ds^2_{S^2}+ \frac{\pi^2}{N_5^2}d\eta^2 \right] + \frac{2 \pi}{N_5} d\rho^2.
\e
In the parentheses we have an $S^3$ with a constant string frame radius proportional to $\sqrt{N_5}$.
The remaining supergravity fields behave as follows
\be
\begin{array}{ll}
e^{\phi} \rightarrow  \frac{\sqrt{\pi}}{2N \sin\left(\frac{\pi }{N_5}\right)} \rho^{1/2} e^{\frac{ \pi \rho}{N_5}}, &
B_{(2)} \rightarrow 2 \left[\frac{N_5}{\pi}\sin\left(\frac{\pi \eta }{N_5}\right)\cos\left(\frac{\pi \eta }{N_5}\right)-\eta \right]  d \Omega_2,\\
& \\
A_{(1)} \rightarrow 0, & A_{(3)} \rightarrow 0.
\end{array}
\e
In all our solutions the 2-sphere vanishes both at $\eta=0$ and at $\eta=\eta_c$, so the
2-sphere times the interval $\eta \in [0,\eta_c]$ is topologically a 3-sphere.
Only the second term in $B_{(2)}$ contributes to the 3-form flux on this 3-sphere, which is given by
\be
\mathcal H_{(3)}=\int_{\tilde S^3} H_{(3)} = - (2 \pi)^2 N_5.
\e

The form of the supergravity fields near $\rho=\infty$ is very similar to the solution for $N_5$
NS5-branes, wrapped on the $AdS_5 \times S^1$ at $\rho=\infty$; the only difference (asymptotically) from the standard linear dilaton solution describing NS5-branes in flat space is some powers of $\rho$ appearing in the expressions for the dilaton and the metric. These powers are related to the fact that the NS5-branes live on a curved space (they also appear in solutions for NS5-branes wrapping spheres, as in \cite{Maldacena:2000yy,Maldacena:2001pb}). For large $N_5$ the type IIA metric is weakly curved near $\rho=\infty$, but the dilaton diverges there, so for large enough $\rho$ we need to lift the solutions again to M theory. Once we lift to M theory, the NS5-branes become M5-branes, and in order to get a sensible solution, we need to specify the positions of these M5-branes on the M theory circle; for instance, if they are all overlapping, our solution will contain a weakly curved region looking approximately like $AdS_7\times S^4$, describing the near-horizon limit of $N_5$ M5-branes on $AdS_5\times S^1$, while if the M5-branes are separated we will have large curvatures (in M theory) in this large $\rho$ regime. The nice thing is that the solution in the region where type IIA string theory is weakly coupled is almost independent of these specific positions, which is why the smearing we performed to go down to type IIA is justified (at least in the region where type IIA string theory is reliable).

\subsubsection{Near the kinks}

 We now analyze the behavior of the various supergravity fields near the points $(\rho=0,\eta=n_i)$,   $i=1,\cdots,L$, where $\lambda'(\eta)$ jumps. Using polar coordinates $\rho=r \cos (\theta),~\eta=n_i+r\sin (\theta)$, $-\pi/2 \leq \theta \leq \pi/2$, we find as $r\to 0$
\be
\begin{array}{ll}
\dot V = \lambda(n_i) + O(r), \quad & V'' = - \frac{1}{2} (a_i-a_{i-1}) \frac{1}{r} + O(r^0), \\
& \\
\dot V' = \frac{1}{2} (a_i-a_{i-1})  (\sin (\theta)-1) - \sum_{j=i+1}^{L+1} (a_j-a_{j-1})+ O(r), \quad & \tilde \Delta = - \lambda(n_i)(a_i-a_{i-1}) \frac{1}{r}.
\end{array}
\e
The metric then behaves as $r\to 0$ as
\be
ds_{10}^2 \rightarrow g(r)^{1/2}(4 ds^2_{AdS_5}+ds^2_{S^2})+g(r)^{-1/2}(dr^2+r^2ds^2_{\tilde S^2}),
\e
where the ${\tilde S^2}$ comes from the $\chi$ and $\theta$ coordinates, and
\be
g(r) = - 4 \frac{\lambda(n_i)}{(a_i-a_{i-1})} r.
\e
The remaining supergravity fields behave as follows near $r=0$ :
\be
\begin{array}{ll}
e^{4\phi} \rightarrow - 64 \frac{1}{\lambda(n_i)(a_i-a_{i-1})^3}r^3, \qquad & B_{(2)} \rightarrow -2 n_i d \Omega_2,\\
& \\
A_{(1)} \rightarrow \left[\frac{1}{2} (a_i-a_{i-1})  (\sin (\theta)-1) - \sum_{j=i+1}^{L+1} (a_j-a_{j-1})\right] d\chi, \qquad & A_{(3)} \rightarrow -2 \lambda(n_i) d\chi \wedge d\Omega_2.
\end{array}
\e
Since the $S^1$ labeled by $\chi$ vanishes at $\rho=0$ both below and above the singular point,
we can construct a two-cycle by joining this $S^1$ with the angular coordinate $\theta$ around
the singularity.
We then have the following non-zero 2-form flux on this two-cycle :
\be
\begin{split}
\mathcal F_{(2)}^i= & \int_{\tilde S^2_i}F_{(2)}^i=  2\pi (a_i-a_{i-1}).\\
\end{split}
\e
This behavior suggests an interpretation of the supergravity
solution as including a stack of $(a_{i-1}-a_{i})$ D6-branes, wrapping an $AdS_5 \times S^2$ space at $\rho=0, \eta = n_i$. One can check that the behavior of the metric and dilaton is consistent with this
interpretation; of course the curvature blows up as $r\to 0$, but that is what we expect from the back-reaction of the D6-branes. In some cases it may be possible to describe the solution in this region as involving D6-brane probes; we will discuss in more detail below when this is reliable. The total number of D6-branes is given by
\be
N_{D6} = \sum_{i=1}^{L}(a_{i-1}-a_{i}) = a_0 - a_L.
\e

\subsubsection{Near the corners of $\Sigma'$}

Finally, our two dimensional space has corners at $\rho=0,~\eta=0, N_5$, and we want to analyze what the solutions look like there.
For (say) $\eta=0$, using polar coordinates $\eta=r \sin (\theta),~\rho=r \cos (\theta)$, with $ 0 \le \theta \le \pi/2$, we get for $r \rightarrow 0$
\be
\begin{array}{ll}
\dot{V} = - \sum_{i=1}^{L+1}(a_i-a_{i-1}) r \sin (\theta) + O(r^2), \qquad & V'' = P(N_5) r \sin (\theta) + O(r^2),\\
& \\
\dot{V}' =  a_0 + O(r), \qquad & \tilde{\Delta} =  \left(\sum_{i=1}^{L+1}(a_i-a_{i-1})\right)^2 + O(r),\\
\end{array}
\e
where we defined
\be
P(N_5)= -\frac{1}{4N_5^2} \sum_{i=1}^{L+1}(a_i-a_{i-1}) \left[ 2 \psi^{(1)}\left(\frac{n_i}{2N_5}\right) - \pi^2 \csc^{2}\left(\pi \frac{n_i}{2N_5}\right) \right].
\e
The result is written in terms of the trigamma function $\psi^{(1)}(x)$, defined in terms of Euler's gamma function by $\psi^{(1)}(x)\equiv  \frac{d^2}{dx^2}\ln  (\Gamma(x))$.
Note that
\be
- \sum_{i=1}^{L+1}(a_i-a_{i-1}) = N_{D6}+a_L = a_0.
\e
The metric has the asymptotic form as $r\to 0$
\be
ds_{10}^2 = \left(\frac{2 a_0}{P(N_5)}\right)^{\frac{1}{2}} \left(4 ds^2_{AdS_5}+ \frac{2P(N_5)}{a_0}[dr^2+ r^2 (d\theta^2+ \cos^2 (\theta) d\chi^2 + \sin^2 (\theta) ds^2_{S^2})]\right),
\e
which looks just like $AdS_5$ times a smooth five dimensional space (written in polar coordinates).
The remaining supergravity fields behave as follows
\be \label{eq:fields.corners}
\begin{array}{ll}
e^{4\phi} \rightarrow  \frac{32}{a_0^3 P(N_5)}, \qquad & B_{(2)} \rightarrow  0,\\
& \\
A_{(1)} \rightarrow  a_0 d\chi, \qquad  & A_{(3)} \rightarrow  - 4 P(N_5) \sin^3 (\theta) r^3 d\chi \wedge d\Omega_2,
\end{array}
\e
and they are also all smooth as $r\to 0$ (with no fluxes localized there).
Thus, we conclude that the apparent singularities at the corners are just artifacts of our choice of coordinates.

A similar analysis shows that the solution is smooth at $\rho=0,\eta=N_5$. In particular one finds that the 1-form potential there is
\be \label{eq:1-form.N_5}
A_{(1)} \rightarrow  a_L d\chi,
\e
as expected since all the D6-branes are in the interval $\rho=0, 0 < \eta < N_5$.

Thus, the only singularities in our space correspond to NS5-branes and D6-branes, and these branes
are exactly the branes we expect to find by taking the near-horizon limit of the D4-branes in the
brane configurations that we started from.


\subsubsection{The 4-form flux}

The supergravity solutions include non-trivial profiles for the NS-NS 2-form potential $B_{(2)}$ and the RR 1-form potential $A_{(1)}$. In such a case, the type IIA 4-form $F_{(4)}$ is not conserved, but rather satisfies $d F_{(4)} =  H_{(3)} \wedge F_{(2)}$. Thus, if we want a
conserved 4-form that could lead to a conserved charge, we need to take a different 4-form
\be \label{tildefour}
{\tilde F}_{(4)} = F_{(4)} + a A_{(1)} \wedge H_{(3)} - \left(1 -a \right) B_{(2)} \wedge F_{(2)},
\e
for some real constant $a$. Generally the Page charge coming from such a 4-form (with $a=0$ or $a=1$) is the only conserved and
quantized charge, but it is not gauge-invariant due to the gauge freedom of shifting $B_{(2)}$ and $A_{(1)}$ (see \cite{Marolf:2000cb} for a general discussion, and \cite{Aharony:2011yc} for a detailed discussion of an analogous situation). In our solutions we can sometimes fix this freedom by requiring ${\tilde F}_{(4)}$ to be non-singular, even when the 1-cycle $S^1$ and the 2-cycle $S^2$ shrink to zero size.

However, it is not possible to find a conserved and globally well-defined 4-form when there are both NS5-branes and D6-branes. The technical issue is that to define a conserved 4-form we need to have either $B_{(2)}$ or $A_{(1)}$ non-singular. However, $A_{(1)}$ jumps by the number of D6-branes as we go along the $\eta$-axis from one side of the D6-brane stack to the other, so it cannot be taken to vanish all along the region where the corresponding 1-cycle vanishes. The same is true for $B_{(2)}$ at $\eta=0$ and $\eta=N_5$, the two opposite sides of the location of the stack of NS5-branes at $\rho=\infty$. The fact that the definition of the D4-brane charge in this case is problematic is related to the fact that \cite{Hanany:1996ie} configurations of D6-branes intersecting NS5-branes carry D4-brane charge (due to the Chern-Simons term in the type IIA supergravity action); and, related to this, the number of D4-branes ending on an NS5-brane (D6-brane) changes as this brane is moved past a D6-brane (NS5-brane), so it is not clear how to identify this number. This is related to our discussion of the linking numbers in section \ref{sec:field.theory}.

\begin{figure}[h]
  \centering
  \includegraphics[scale=.7]{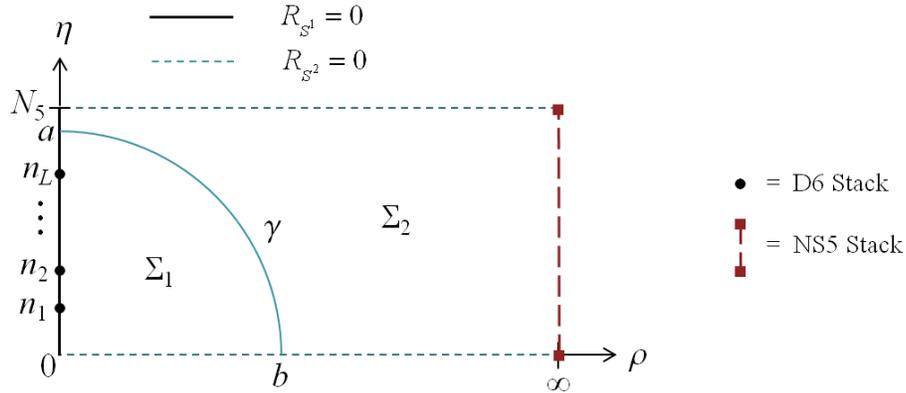}
      \caption{The Riemann surface $\Sigma'$. We depict the first choice for the surface $\Sigma_1$ on which ${\cal F}_1$ is well-defined, and for the surface $\Sigma_2$ on which ${\cal F}_2$ is well-defined, separated by the curve $\gamma$.}
      \label{fig:patches}
\end{figure}

Nevertheless, there is a natural way to define a conserved 4-form charge in our solutions. The 4-form ${\cal F}_1 \equiv F_{(4)} - B_{(2)} \wedge F_{(2)}$ is well-defined near all the stacks of D6-branes at $\rho=0,\eta=n_i$, and the 4-form ${\cal F}_2 \equiv F_{(4)} + A_{(1)} \wedge H_{(3)}$ is well-defined near the stack of NS5-branes at $\rho=\infty$. We can extend the regions where these two 4-forms are well-defined so that together they cover all of $\Sigma'$. The main constraint is that the region $\Sigma_2$ where
${\cal F}_2$ is well-defined cannot include any point in the interval $\rho=0,\eta \in [n_1,n_L]$ (where the $S^1$ on which $A_{(1)} \neq 0$ shrinks to zero size), while the region $\Sigma_1$ where ${\cal F}_1$ is well-defined cannot include the point $\rho=\infty$. This leaves us with two possible choices for the topology of these regions. We can take $\Sigma_1$ to be a region that intersects the $\eta$-axis
along $[0,a]$ and the $\rho$-axis along $[0,b]$, where $n_L < a < N_5$ and $b<\infty$, and $\Sigma_2$ to be the complement
of this region, see figure \ref{fig:patches}; this fulfills the requirements above. There is then a unique non-singular choice for ${\cal F}_1$ in $\Sigma_1$, given by choosing $A_{(1)}$ (by a large gauge transformation) to vanish on $\rho=0,\eta \in [n_L,N_5]$, and similarly there is a unique
non-singular choice for ${\cal F}_2$ in $\Sigma_2$ by choosing $B_{(2)}$ (by a large gauge transformation) to vanish at $\eta=0$. The other choice is to take $\Sigma_1$ to be
a region that intersects the $\eta$-axis along $[{\tilde a}, N_5]$ and covers the interval $\rho \in [0,\tilde b],\eta=N_5$ with $0<\tilde a < n_1$ and ${\tilde b}< \infty$. The two choices are related by $\eta \rightarrow N_5 - \eta$, so we will focus on the first choice here.

At first sight, the fact that we used two different 4-forms to cover $\Sigma'$ does not allow us to
obtain a conserved charge. However, consider the integral of ${\cal F}_1$ on the boundary $\partial \Sigma_1$ of $\Sigma_1$ (times $S^1 \times S^2$). Since $d{\cal F}_1=0$, this integral vanishes. On the other hand, it has two contributions; one from the ``external'' boundary of $\partial \Sigma_1$ which is along $\partial \Sigma'$, where it gets contributions from the stacks of D6-branes (and not from any other points on the boundary), and one from the ``internal'' boundary, along the curve $\gamma$ in figure \ref{fig:patches}. Similarly, the integral of ${\cal F}_2$ on $\partial \Sigma_2$ also vanishes, and it is given by the contribution from the stack of NS5-branes at $\rho=\infty$, plus the contribution from the ``internal'' boundary. If we add these two integrals, the total contribution from the ``internal boundary'' $\gamma$ is the integral of
\be
{\cal F}_1-{\cal F}_2 =  - d(B_{(2)}\wedge A_{(1)})
\e
along this boundary; but this is just proportional to the difference in the values of $B_{(2)}\wedge A_{(1)}$ between the two edges of this boundary $\gamma$ at $\rho=0,\eta=a$ and $\rho=b,\eta=0$, and this vanishes since either $B_{(2)}$ or $A_{(1)}$ vanishes at each of these points. Thus, we find that the sum of the 4-form fluxes ${\cal F}_1$ or ${\cal F}_2$ over all the stacks of D6-branes and the stack of NS5-branes vanishes, so this
defines a conserved charge. It is natural to relate this charge to the number of D4-branes ending
on the D6-branes or the NS5-branes, so we identify it with the linking number discussed in section \ref{sec:field.theory}.

Let us now apply the condition above to fix $A_{(1)}$ and $B_{(2)}$. We must impose that $A_{(1)}$ vanishes on the interval $\rho=0,\eta \in [n_L,N_5]$ and $B_{(2)}$ on the $\rho$-axis, by judiciously making large gauge transformations. In our solutions $B_{(2)}$ already vanishes on the $\rho$-axis, while to achieve the condition on $A_{(1)}$, in view of equation \eqref{eq:1-form.N_5}, we perform the large gauge transformation
\be
A_{(1)} \rightarrow A_{(1)} - a_L =  \left(\frac{2\dot V \dot V'}{2 \dot V - \ddot V}-a_L \right) d \chi.
\e
With this choice of RR 1-form potential, the 4-form flux coming from the stack of NS5-branes is
\be
\int_{\Sigma_4} \mathcal F_2 = -(2\pi)^3 a_L N_5,
\e
and the 4-form flux coming from the $i^{th}$ stack of D6-branes is
\be
\int_{\Sigma_4^i} \mathcal F_1 = (2\pi)^3 n_i (a_i-a_{i-1}).
\e
The sums of all these fluxes vanish, as can be seen by using the constraint equation \eqref{eq:lambda.const}.

\subsection{The dual gauge theory}

Type IIA string theory on the spacetimes described above was claimed
in \cite{Gaiotto:2009gz} to be dual to the $\mathcal N=2$ superconformal field theories reviewed in section \ref{sec:field.theory}.
We can make this translation precise
by using the brane interpretation. The general solution \eqref{eq:gen.sol} is dual to a system of D4-branes stretched between and intersecting a stack of $N_5$ NS5-branes and $L$ stacks of D6-branes, in the near-horizon of the D4-branes. We can read off the numbers of D6-branes in each stack from the
2-form flux, and their linking numbers from the 4-form flux, as described in the previous subsection. As described in section \ref{sec:field.theory}, this is
enough to identify the dual gauge theory. We find that :

\begin{itemize}

\item There is an $SU(\lambda_n)$ gauge group factor for each integer value $n=1,2,...,N_5-1$ of $\eta$, where $\lambda_n = \lambda(n)$.

\item If there is a kink at $\eta=n$, then there are an extra $(a_{n-1}-a_n)$ fundamental hypermultiplets charged under $SU(\lambda_n)$.

\item The supergravity solution describes the strong coupling limit of the gauge theory, when all the gauge couplings $g_n$ become arbitrarily strong.

\end{itemize}

The last observation follows from the fact that in the solutions we described up to now there is a single stack of NS5-branes (so all the NS5-branes in the brane configuration are on top of each other). We can think of the $2L$ parameters $\{k_i,n_i\}$ of the solution as describing the number of D6-branes in each stack and the linking number of the D6-branes in each stack, respectively.

 In the classical analysis of the brane system of section \ref{sec:field.theory} it is natural to separate the NS5-branes into multiple stacks, making some gauge couplings finite (while others are still infinite).
In the next section we construct the supergravity duals to a system with an arbitrary number of stacks of NS5-branes. In this way we establish a one-to-one correspondence between our supergravity solutions and {\it all} possible strong coupling limits of the associated $\mathcal N=2$ SCFTs.

\subsection{Examples of supergravity solutions}

\begin{figure}[h]
  \centering
  \includegraphics[scale=.7]{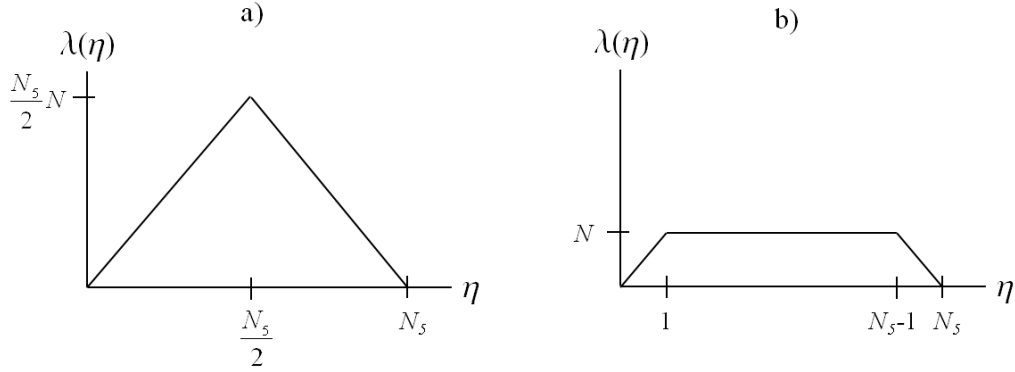}
      \caption{a) Simple example of a line charge density $\lambda(\eta)$ describing one stack of $2N$ D6-branes. b) Simple example of a line charge density $\lambda(\eta)$ describing two stacks of $N$ D6-branes.}
      \label{fig:simple.lambda}
\end{figure}

We now provide the explicit solutions dual to the specific SCFTs analyzed in subsection \ref{subsec:Examples}, when all the gauge couplings $g_n$ become arbitrarily strong. The simplest solutions are described by line charge densities with a single kink (one stack of D6-branes). A simple example of such a line charge density $\lambda(\eta)$, invariant under $\eta \rightarrow N_5-\eta$, has the following profile
\be \label{eq:simple.lambda0}
\lambda(\eta)= \left\{ \begin{array}{ccrcl}
                       N\eta & \textrm{if} & 0 \leq & \eta & \leq N_5/2 \\
                       N(N_5-\eta) & \textrm{if} & N_5/2 \leq & \eta & \leq N_5.  \\
                       \end{array}
\right.
\e
It is depicted in figure \ref{fig:simple.lambda} a). The associated solution is
\be \label{eq:Vdot}
\begin{split}
\dot V(\rho,\eta)= -\frac{1}{2} \sum_{m=-\infty}^{\infty} & \left[2 N \left( \sqrt{ \rho^2 + (\eta - 2mN_5 - N_5/2)^2} - \sqrt{\rho^2 + (\eta - 2mN_5 + N_5/2)^2 } \right) - \right. \\
& \left.   - N  \left( \sqrt{ \rho^2 + (\eta - 2mN_5 - N_5)^2} - \sqrt{\rho^2 + (\eta - 2mN_5 + N_5)^2 } \right) \right].
\end{split}
\e
The corresponding supergravity solution has one stack of $2N$ D6-branes and one stack of $N_5$ NS5-branes. It is dual to the linear quiver in figure \ref{fig:conf.quiver} a), when all the gauge couplings are taken to infinity.

Another example, that was analyzed in detail in \cite{ReidEdwards:2010qs}, is the solution described by a simple line charge density with two kinks. This line charge density $\lambda(\eta)$, as depicted in figure \ref{fig:simple.lambda} b), has the following profile invariant under $\eta \rightarrow N_5-\eta$ :
\be \label{eq:simple.lambda}
\lambda(\eta)= \left\{ \begin{array}{ccrcl}
                       N\eta & \textrm{if} & 0 \leq & \eta & \leq 1 \\
                       N     & \textrm{if} & 1 \leq & \eta & \leq N_5-1 \\
                       N(N_5-\eta) & \textrm{if} & N_5-1 \leq & \eta & \leq N_5.  \\
                       \end{array}
\right.
\e
With this choice of parameters its periodic extension can be written in a compact form
\be
\lambda_p(\eta) = \frac{N}{2} \sum_{m=-\infty}^{\infty} \sum_{i=1}^3 \left( |\eta + 2mN_5 + \nu_i| - | \eta + 2mN_5 - \nu_i| \right) ,
\e
where $\nu_1=n_1=1, \nu_2=n_2=N_5-1, \nu_3=-n_3=-N_5$. The solution is
\be \label{eq:Vdot}
\begin{split}
\dot V(\rho,\eta)= & -\frac{N}{2} \sum_{m=-\infty}^{\infty} \sum_{i=1}^3 \left( \sqrt{ \rho^2 + (\eta - 2mN_5 - \nu_i)^2} - \sqrt{\rho^2 + (\eta - 2mN_5 + \nu_i)^2 } \right). \\
\end{split}
\e
The supergravity solution has two stacks of $N$ D6-branes and one stack of $N_5$ NS5-branes. It is dual to the linear quiver in figure \ref{fig:conf.quiver} b), when all the gauge couplings are taken to infinity.

\subsection{Validity of the supergravity solution}
\label{validity}

In this subsection we discuss for which values of the parameters the solutions above provide a good
description of the holographic dual of the field theory. Apriori we require that the solutions should
have weak coupling (otherwise we need to lift to M theory), and small curvatures (otherwise we need
to take into account stringy corrections to supergravity). However, we saw that (as we expected on
general grounds) all our solutions involve regions which look like D6-branes (where the curvature
diverges) and like NS5-branes (where the string coupling diverges). So, we cannot require small
couplings and curvatures everywhere. The minimal thing we can try to require is to have small couplings
and curvatures far away from the D6-branes and NS5-branes (this requires in particular that the radius
of the region where their back-reaction is large is smaller than the typical scale in the geometry).
In such a case we could reliably describe all the bulk fields, but not the fields living on the
branes (which include, in particular, gauge fields living on the D6-branes giving the flavor symmetry
currents). We could also ask whether the low-energy fields living on the D6-branes and NS5-branes are
weakly coupled or not, and when they are we could describe these fields reliably as well. Finally, in some cases it may be possible to reliably treat the D6-branes as probes, in which case we could reliably describe all the open string fields living on the D6-branes (with small corrections from open
string loops).

A first obvious comment is that near the NS5-branes the solution is weakly curved (in the string frame) only for large $N_5$, so this is a necessary condition for the validity of our solutions. Similarly, if
we lift the solution near the NS5-branes to M theory, its curvature (in eleven dimensional Planck
units) again goes as a negative power of $N_5$. In particular
this means that the solution for $N_5=2$ (corresponding to the $SU(N_c)$ theory with $2N_c$ flavors) is not reliable, and we need to go beyond supergravity in order to construct the holographic dual of this field theory (the dual of this specific theory was recently discussed in \cite{Gadde:2009dj,Gadde:2010zi}). It is possible that, as for NS5-branes in flat space, one can find an exact worldsheet theory that describes the region close to the NS5-branes, and that this region could then be controlled even for small values of $N_5$ (at least in the string theory regime where we do not lift to M theory) -- this is a topic that we leave to future investigations.

Next, let us analyze various scalings we can perform in our solutions, related to the scalings of field
theory parameters that we discussed in section \ref{sec:field.theory}. One thing we can do is
to multiply $V$ (and thus also $\lambda(\eta)$) by some integer $K_1 > 1$. This leaves the string
frame metric intact, while multiplying the 2-form and 4-form fluxes by $K_1$ (but not the 3-form
flux), and multiplying $1/g_s$ by $K_1$ as well. In the field theory this is identified with
scaling the numbers of D4-branes and D6-branes, but not of NS5-branes; $N$ is taken to be large
with fixed $g_s N$ and with a fixed quiver diagram. This limit decreases the closed string loop
corrections to our solutions,
while keeping the curvature corrections and open string loop corrections the same.
This limit is similar to the usual 't Hooft large $N$ limit.

Another scaling we can perform is to take $\eta \to K_2*\eta$, for some integer $K_2$, and to take
$V(\rho,\eta)$ to ${\tilde V}(\rho,\eta) = K_2 V(\rho/K_2,\eta/K_2)$ which is still a solution to the
equations of motion. This transformation multiplies $N_5$ and the numbers of D4-branes by $K_2$, while keeping the numbers of D6-branes (related to the $a_i$) fixed. The distances in the string frame
(for instance, the radius of the $S^2$) become larger by a factor of $\sqrt{K_2}$ (thus decreasing
the string frame curvature by a factor of $K_2$), and the string coupling constant is also
multiplied by $\sqrt{K_2}$. So, this scaling increases the string loop corrections while reducing
the curvature corrections to our solutions. If we perform this scaling together with the one of the
previous paragraph (with $K_1=K_2$), then the numbers of NS5-branes and D6-branes are both rescaled,
and both string loop and curvature corrections become smaller. However, the open string loop corrections
on the D6-branes become larger in this rescaling.

At low energies, the theory on the $i^{th}$ stack of D6-branes is a $U(a_{i-1}-a_i)$ gauge theory living on $AdS_5\times S^2$, with coupling constant $g_{YM,D6}^2 = g_s (\alpha')^{3/2}$ (up to constants of order one). The AdS/CFT correspondence maps this to the flavor symmetry of the relevant hypermultiplets in the
fundamental representation. This gauge theory gives a good description of the low-energy physics on the D6-branes in our solutions as long as the 't Hooft coupling constant $g_{YM,D6}^2 (a_{i-1}-a_i)$ is small at the scale of the $AdS_5\times S^2$ that the $6+1$ dimensional field theory lives on. In our solutions the size of the
$AdS_5\times S^2$ and the dilaton go to zero near the D6-branes, but this is because of the back-reaction of the D6-branes themselves, and by comparing the solution near the D6-brane to the general near-horizon
limit of D6-branes we can extract the gauge coupling constant on the D6-branes at the scale of the space that the D6-brane lives on. We find that this goes as $g_{YM,D6}^2 (a_{i-1}-a_i) / R_{AdS_5}^3 =
(a_{i-1}-a_i) / \lambda(n_i)$, up to numerical constants. Thus, the low-energy gauge theory on the D6-branes is weakly coupled as long as $(a_{i-1}-a_i) \ll \lambda(n_i)$ for all $i$. Generally this will
be true in our solutions as long as the $n_i$'s are large, namely we have a product of a large number
of gauge groups. Note that this is not true in the large $N$ limit of the two specific examples discussed
above, so in these examples the theory on the D6-branes is strongly coupled.

Similarly, at low energies, the theory on the NS5-branes is a six dimensional $(2,0)$ superconformal field theory living on $AdS_5\times S^1$; the dimensional reduction of this theory on the $S^1$ leads to a
low-energy theory which is a $U(N_5)$ gauge theory living on $AdS_5$, with coupling constant $g_{YM,NS5}^2 = R_{S^1}$ (up to constants of order one). This gauge theory maps to an enhanced global symmetry that the
field theories reviewed in section \ref{sec:field.theory} have at infinite coupling (when all the NS5-branes in the
brane construction overlap). This gauge theory gives a good description of the low-energy physics on the NS5-branes in our solutions, at scales below $1/R_{S^1}$, as long as the coupling constant $g_{YM,NS5}^2 N_5$ is small at the scale of the $AdS_5$ that the $4+1$ dimensional SYM theory
lives on. Since in our solutions near the NS5-branes $R_{AdS_5} = R_{S^1}$, this is never true, so we
cannot approximate the low-energy theory living on the NS5-branes by a Yang-Mills theory.

\section{Solutions with multiple stacks of NS5-branes}

The geometry and topology of the solutions of the previous section suggest a natural generalization of the boundary conditions \eqref{eq:b.c.}, to construct solutions with $m>1$ stacks of NS5-branes. The form of the metric \eqref{eq:10d.metric} implies that a stack of NS5-branes can only be located at $\rho=\infty$. Furthermore, the appearance of a 3-cycle due to the boundary conditions at $\eta=0,N_5$ indicates that, to obtain a geometry with $m$ 3-cycles, it is necessary to impose similar boundary conditions ${\dot V}=0$ at $m-1$ lines located at constant values $\eta=\eta_k, k=1,...,m-1$, with $0 < \eta_1<\eta_2<...<\eta_{m-1} < N_5$ (quantization of the 3-form flux will imply that the $\eta_k$ should be integers). These lines must extend to $\rho=\infty$ where the 5-branes sit, and they must be parallel to the $\rho$ axis, but they do not need to extend all the way to $\rho=0$; indeed they cannot do this without violating our boundary condition for $\dot V$ there.
Thus, such a solution should be subject to the following $m-1$ extra boundary conditions
\be \label{eq:b.c.new}
\dot V|_{\rho \in [L_1,\infty),~ \eta = \eta_1}=\dot V|_{\rho \in [L_2,\infty),~ \eta = \eta_2}=...=\dot V|_{\rho \in [L_{m-1},\infty),~ \eta = \eta_{m-1}}=0,
\e
with some parameters $L_k > 0$. These conditions can be seen as introducing $m-1$ semi-infinite cuts to the half-strip. If we think of the problem in electrostatic terms, the boundary conditions \eqref{eq:b.c.new} correspond to $m-1$ {\it infinite conducting planes with concentric circular apertures of radii} $\{L_k\}$ located at $\{\eta=\eta_k\}$ .

\begin{figure}[h]
  \centering
  \includegraphics[scale=1.3]{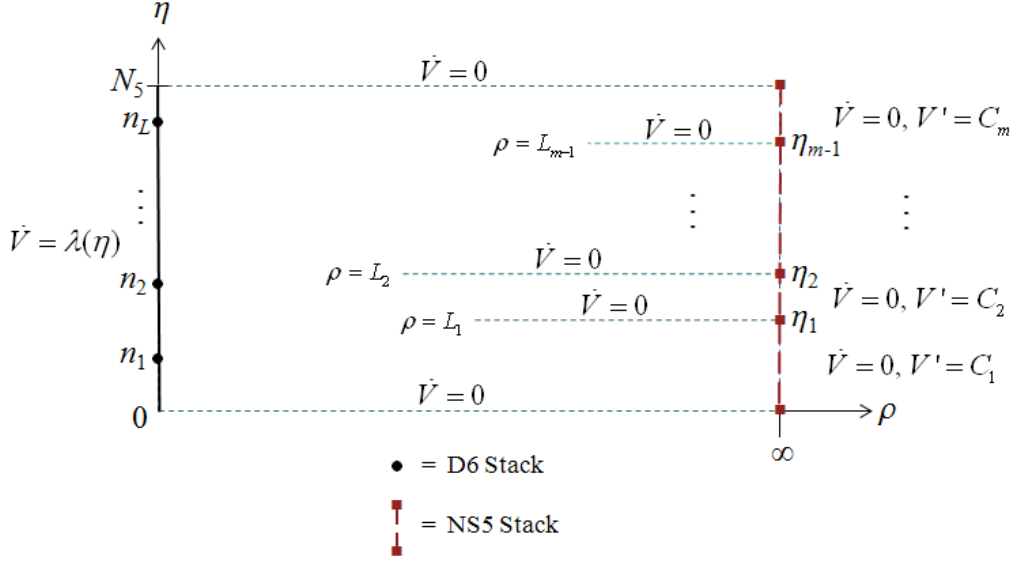}
      \caption{The boundary conditions to the Laplace equation in terms of $\dot V$ when we separate the NS5-branes into $m$ stacks, corresponding to making $(m-1)$ gauge couplings finite. These boundary conditions determine the behavior of $V'$ at $\rho=\infty$, which in turn controls the value of the $(m-1)$ finite gauge couplings.}
      \label{fig:m.stacks}
\end{figure}

Apriori it is not clear if this is sufficient to separate the NS5-branes into multiple stacks, since the $\eta$-coordinate shrinks to zero size at $\rho=\infty$.
In order to understand this better, we consider the original coordinates $(r,y)$; in these coordinates
separating the 5-brane stacks means putting them at different values of $r$ (while we still smear them over
the angular coordinate $\beta$). Recall that $V'=\ln(r)$, and note that
we have the freedom to shift $V'$ by an overall constant without changing the solution. For the solution with one stack of NS5-branes, we have $V' \xrightarrow{\rho \rightarrow \infty} C$ where $C$ is an arbitrary (unphysical) constant parameterizing this freedom. These observations imply that a solution in which the $m$ stacks of NS5-branes are separated in the $r$-direction should have the following boundary values :
%
\be \label{eq:b.c.Vp}
V'|_{\rho=\infty,~\eta \in [0,\eta_1]}=C_1, \quad V'|_{\rho=\infty,~\eta \in [\eta_1,\eta_2]}=C_2, \quad ... \quad ,  V'|_{\rho=\infty,~\eta \in [\eta_{m-1},N_5]}=C_m,
\e
with $\{C_k\}$ being different constants. We will see below that indeed this behavior of $V'$ follows from \eqref{eq:b.c.new}, with some specific relation between the parameters $L_k$ and $C_{k+1}-C_k$. A schematic representation of the boundary conditions for a solution with $m>1$ stacks of NS5-branes is given in figure \ref{fig:m.stacks}. Note the presence of new non-trivial 3-cycles as compared to the solution with one stack of NS5-branes. We will show this explicitly below.

\subsection{General solution with $m>1$ stacks of NS5-branes}

A general solution with $m>1$ stacks of NS5-branes is defined by the line charge density \eqref{eq:gen.lambda} and by $m-1$ cuts parameterized by $2(m-1)$ constants $\{L_k,\eta_k\}$ (see figure \ref{fig:m.stacks}). The constants $\{C_k\}$ are fixed by the latter as we explain below. Unfortunately we were not able to find
the general solution to the Laplace equation with these boundary conditions, but we can easily understand some of its general properties.
The solution restricted to the rectangular region defined by $\rho < \textrm{min}(\{L_k\})$ is subject to the boundary conditions \eqref{eq:b.c.} and takes the form
\be \label{eq:Vd0}
\dot V|_{\rho < \textrm{min}(\{L_k\})} \equiv \dot V_0 = \sum_{n=1}^{\infty} \rho \frac{n \pi}{N_5} \left[ A_{n} K_1 \left(\frac{n \pi}{N_5} \rho\right) + B_{n} I_1 \left(\frac{n \pi}{N_5}\rho\right)\right]\sin \left(\frac{n \pi}{N_5} \eta\right),
\e
where $A_n$ is determined by the boundary condition at $\rho=0$ and is given by the same expression as in the $m=1$ case \eqref{eq:An}. The solution restricted to one of the $m$ semi-infinite strips defined by $\rho > \textrm{max}(L_k,L_{k-1}),\eta \in [\eta_{k-1},\eta_k] $ (with $L_0=L_m=0, \eta_0=0, \eta_m=N_5$) is subject to the boundary conditions
\be
\dot V|_{\eta=\eta_{k-1},\eta_k}=0, \qquad \dot V \xrightarrow{\rho \rightarrow \infty} 0.
\e
Thus, it can be written as
\be  \label{eq:Vdk}
\dot V|_{\rho > \textrm{max}(L_k,L_{k-1}), \eta \in [\eta_{k-1},\eta_k]} \equiv \dot V_k=\sum_{n=1}^{\infty} \rho \frac{n \pi}{\delta \eta_k} A_{n}^{(k)} K_1 \left(\frac{n \pi}{\delta \eta_k} \rho\right) \sin \left(\frac{n \pi}{\delta \eta_k} (\eta-\eta_{k-1})\right),
\e
where $\delta \eta_k = \eta_k-\eta_{k-1}$. Note that we used the boundary condition at $\rho=\infty$ to fix $B_n^{(k)}=0$.
We also have
\be \label{eq:V'k}
\begin{split}
V'|_{\rho > \textrm{max}(L_k,L_{k-1}),\eta \in [\eta_{k-1},\eta_k]} \equiv V'_k= & - \sum_{n=1}^{\infty} \frac{n \pi}{\delta \eta_k}  A_{n}^{(k)} K_0 \left(\frac{n \pi}{\delta \eta_k} \rho\right) \cos\left(\frac{n \pi}{\delta \eta_k} (\eta-\eta_{k-1})\right)+C_k.
\end{split}
\e
The various integration constants appearing in the solution (e.g. $C_k, B_n, A_n^{(k)}$) are in principle determined in terms of $\{L_k, \eta_k\}$ by requiring continuity of $\dot V$ and $V'$
at the overlaps of the different regions. Note that $V'$ is not continuous along the cuts at $\eta=\eta_k$, but jumps as we go from one side to the other (in the full geometry the two sides of the cut are at
different positions in space-time).

In the rest of this section we use the implicit form of the solution \eqref{eq:Vd0}, \eqref{eq:Vdk} and \eqref{eq:V'k} to clarify the topology of the solution, show that it is smooth and give its precise field theory interpretation.

\subsection{Behavior near special points}

\begin{figure}[h]
  \centering
  \includegraphics[scale=1.3]{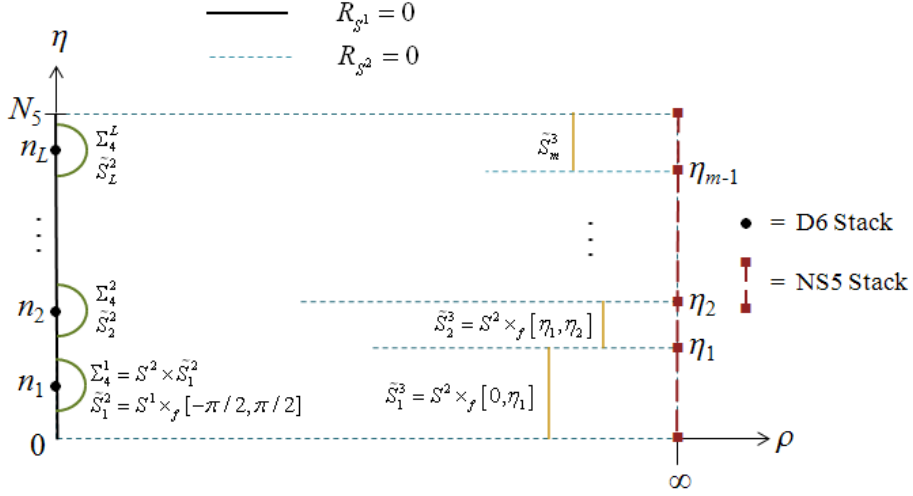}
      \caption{Non-trivial cycles of the solution with $m$ stacks of NS5-branes and $L$ stacks of D6-branes.}
      \label{fig:topology1}
\end{figure}

To understand the topology of the solution, we analyze the behavior of the supergravity fields near the zeros and singular points of the potential $V(\rho,\eta)$ and its various derivatives. One can show that the analysis of the solution \eqref{eq:Vd0} at special points with $\rho=0$ ($\eta=n_i$, $i=1,\cdots,L$ and $\eta=0,N_5$)  remains unchanged with respect to the case of one stack of NS5-branes. Thus, the geometry of the general supergravity solution parameterized by \eqref{eq:Vd0} still contains $L$ non-trivial 2-cycles $\tilde S^2_i$ and 4-cycles $\Sigma_4^i$ as depicted in figure \ref{fig:topology1}. But new features arise due to the cuts, that we analyze in detail below.

\subsubsection{Asymptotic region at $\rho \rightarrow \infty$}

We want the behavior of the various derivatives of the potential $V(\eta,\rho)$ in the asymptotic region $\rho \rightarrow \infty, \eta \in [\eta_{k-1},\eta_k]$ (with $\eta_0=0, \eta_m=N_5$). As shown above, the solution restricted to $\rho > \textrm{max} (L_{k-1},L_k)$ (with $L_{0}=L_m=0$) and $ \eta \in [\eta_{k-1},\eta_k]$ is given by
\be
\dot V_{k}= \sum_{m=1}^{\infty} \rho  \frac{m \pi}{\delta \eta_k}   A_{m}^{(k)} K_1 \left(\frac{m \pi}{\delta \eta_k} \rho \right) \sin\left(\frac{m \pi}{\delta \eta_k} (\eta-\eta_{k-1}) \right),
\e
where $\delta \eta_k = \eta_k - \eta_{k-1}$. Following the analysis of the previous section, the metric has the asymptotic form
\be
ds_{10}^2 \rightarrow 4 \rho(ds^2_{AdS_5}+d\chi^2)+\frac{2 \delta \eta_k}{\pi}\left[\sin^2\left(\frac{\pi (\eta-\eta_{k-1}) }{\delta \eta_k}\right)ds^2_{S^2}+ \frac{\pi^2}{\delta \eta_k^2}d\eta^2 \right] + \frac{2 \pi}{\delta \eta_k} d\rho^2.
\e
In the parentheses we have an $S^3$ with a constant string frame radius proportional to $\sqrt{\delta \eta_k}$.
The remaining supergravity fields behave as follows
\be
\begin{array}{ll}
e^{\phi} \rightarrow  \frac{\sqrt{\pi}}{2N \sin\left(\frac{\pi }{\delta \eta_k}\right)} \rho^{1/2} e^{\frac{ \pi \rho}{\delta \eta_k}}, &
B_{(2)} \rightarrow 2 \left[\frac{\delta \eta_k}{\pi}\sin\left(\frac{\pi (\eta-\eta_{k-1}) }{\delta \eta_k}\right)\cos\left(\frac{\pi (\eta-\eta_{k-1}) }{\delta \eta_k}\right)-\eta \right]  d \Omega_2,\\
& \\
A_{(1)} \rightarrow -a_L d\chi, & A_{(3)} \rightarrow 0.
\end{array}
\e
The 2-sphere vanishes both at $\eta=\eta_{k-1}$ and at $\eta=\eta_k$, so the
2-sphere times the interval $\eta \in [\eta_{k-1},\eta_k]$ is topologically a 3-sphere.
Only the second term in $B_{(2)}$ contributes to the 3-form flux on this 3-sphere :
\be
\mathcal H_{(3)}=\int_{\tilde S^3} H_{(3)} = - (2 \pi)^2 \delta \eta_k.
\e
The form of the supergravity fields signals the presence of a stack of $\delta \eta_k$ NS5-branes localized at $\rho=\infty$ and wrapping an $AdS_5 \times S^1$ subspace. The 4-form flux coming from this stack of NS5-branes is
\be
\int_{\Sigma_4^k} \mathcal F_2 = -(2\pi)^3 a_L \delta \eta_k,
\e
which is identified with the linking number of $\delta \eta_k$ NS5-branes, each with linking number $a_L$, in complete agreement with the classical brane picture (which implied that the linking numbers of all NS5-branes must be equal).

We conclude that the solution describes $m$ stacks of NS5-branes, one for each value of $k$. We also find a non-trivial 3-cycle for each value of $k$ as depicted in figure \ref{fig:topology1}, giving a total of $m$ 3-cycles.
Note that the total number of NS5-branes is $\sum_{k=1}^m \delta \eta_k = N_5$.

\subsubsection{Near the tip of the cuts}

Along the cuts at $\eta=\eta_k$ our solutions are not smooth, and $V'$ jumps as we cross
the cut. However, the full supergravity solution is smooth, as we show next by analyzing
the solution near
the tip of the cuts $\{(\rho=L_k,\eta=\eta_k)\}$. For a given $k$ we introduce complex coordinates
\be
w= \left[ (L_k-\rho) + i \left( \eta - \eta_k \right) \right]^{\frac{1}{2}}.
\e
The Laplace equation for $V(\rho,\eta)$ takes the form
\be
\begin{split}
& [4L_k \partial_w \partial_{\bar w}- \bar w \partial_w - w \partial_{\bar w} - 2 (w^2 + \bar w^2) \partial_w \partial_{\bar w}] V = 0. \\
\end{split}
\e
As $w \rightarrow 0$ it reduces to $\partial_w \partial_{\bar w} V =0$.
So, a real solution has the asymptotic expansion near $w\to 0$
\be
V= -a_k w^{\alpha} - \bar a_k \bar w^{\bar \alpha} + O(w^{\alpha+1}),
\e
for some constants $\alpha$ and $a_k$.
The boundary condition \eqref{eq:b.c.new} at $\rho=L_k,\eta = \eta_k$ becomes $\dot V|_{\textrm{Re}(w)=0} = 0$. It implies $a_k,\alpha \in \mathbb{Re}$ and $ \alpha -2 = n \in \mathbb{Z}^+$. Below we prove that $\dot V / V'' > 0$ and $V''>0$ everywhere in the interior of $\Sigma'$. The solution has to satisfy these inequalities also near the cut; they lead to $n=1$ and $a_k > 0$, respectively.
Then, in terms of polar coordinates $w^2 = r e^{-i(\theta-\pi)}$ (for which $\theta=0$ on one side of the cut, and $\theta=2\pi$ on the other side) we get for $r \rightarrow 0$
\be
\begin{array}{ll}
\dot V =  3 a_k L_k r^{1/2} \sin \left(\frac{\theta}{2} \right) + O(r), \quad  & V'' = \frac{3}{2} a_k \frac{1}{r^{1/2}} \sin \left(\frac{\theta}{2} \right) + O(r^0), \\
& \\
\dot V' =  \frac{3}{2} a_k L_k \frac{1}{r^{1/2}} \cos \left(\frac{\theta}{2} \right) + O(r^0),\quad  & \tilde \Delta = \frac{9}{4} a_k^2 L_k^2  \frac{1}{r} + O\left(\frac{1}{r^{1/2}}\right). \\
\end{array}
\e
The metric then has the asymptotic form
\be
ds_{10}^2 \rightarrow L_k \left[ 4 ds^2_{AdS_5} + 4 \frac{1}{L_k} r \sin^2 \left(\frac{\theta}{2} \right) ds^2_{S^2} + \frac{1}{L_k} \frac{1}{r} (dr^2  + r^2 d\theta^2) + 4 d\chi^2 \right].
\e
which looks just like $AdS_5$ times a smooth five dimensional space $\mathbb{R}^4\times S^1$ (written in polar coordinates).
The remaining supergravity fields behave as follows
\be \label{eq:fields.tip}
\begin{array}{llll}
e^{\phi} \rightarrow  \frac{2}{3} \frac{1}{a_k}, \quad & B_{(2)} \rightarrow  -2 \eta_k d\Omega_2, \quad & A_{(1)} \rightarrow  0, \quad  & A_{(3)} \rightarrow  0,\\
\end{array}
\e
and they are also all smooth as $r\to 0$ (with no non-trivial fluxes localized there).
Thus, we conclude that the supergravity solution is smooth at the tip of the cuts, despite the discontinuity in $V'$ and $\dot V'$ along the cut.

\subsubsection{Positivity of the metric}

Finally, we need to verify that our general solutions satisfy the positivity condition \eqref{eq:metric+condition} everywhere. The proof that $\dot V>0$ goes along the same lines as for the case of one stack of NS5-branes, by using Hopf's maximum principle \cite{Hopf:1927} and the boundary conditions \eqref{eq:b.c.Vdot} and \eqref{eq:b.c.new}.

To prove that $V''>0$ we define a new function $\tilde V''=V''-V''_{1s}$, where $V''_{1s}$ is the solution with one stack of NS5-branes \eqref{eq:gen.sol}. Since both $V''$ and $V''_{1s}$ satisfy the axially symmetric Poisson equation in three dimensions \eqref{eq:Poisson}, it follows that $\tilde V''$ satisfies the axially symmetric Laplace equation in three dimensions \eqref{eq:Laplace}. The boundary conditions for $\tilde V''$ are then derived from those of $V''$ and $V''_{1s}$. We find (for $\left[ (L_k-\rho) + i ( \eta - \eta_k ) \right]^{\frac{1}{2}}=r e^{-i(\theta-\pi)}$ as above)
\be
\begin{array}{ll}
\tilde V''|_{\eta=0,N_5}=0, \quad & \tilde V'' \xrightarrow{\rho \rightarrow \infty} 0,\\
& \\
\tilde V''|_{\rho>L_k,\eta=\eta_k} = f(\rho) \delta(\eta-\eta_k) + \text{finite}, \quad & \tilde V'' \xrightarrow{r \rightarrow 0} \frac{3}{2} a_k \frac{1}{r^{1/2}} \sin \left(\frac{\theta}{2} \right) + \textrm{finite},\\
\end{array}
\e
where $f(\rho)$ is non-negative, and obeys
\be
f(\rho) \sim (\rho-L_k)^{1/2} \quad \textrm{as} \quad \rho \rightarrow L_k, \qquad f(\rho) \xrightarrow{\rho \rightarrow \infty} C_{k+1}-C_k.
\e
The maximum principle for the Laplace equation \cite{Hopf:1927} then implies that $\tilde V''>0$ in the interior. Since we proved in section 4 that  $V''_{1s}>0$ in the interior, it then follows that also $V''>0$. This establishes positivity of the metric.

\subsection{Field theory interpretation}

We conclude that the supergravity solutions described in this section are dual to D4-branes stretched between and intersecting $L$ stacks of D6-branes and $m$ stacks of NS5-branes, in the near-horizon limit of the D4-branes. They describe the strong coupling limit of the corresponding linear quiver (see section \ref{sec:field.theory}), when the $m-1$ gauge couplings $\{g_{\eta_{k}}\}$ of the gauge groups coming from D4-branes stretched between the $m$ stacks of NS5-branes are kept fixed, and all other gauge couplings are taken to infinity. The $2L$ parameters $\{a_i-a_{i-1},n_i\}$ give the number of D6-branes in each stack and the linking number of each stack, respectively. Similarly, the $2(m-1)$ independent parameters $\{\delta \eta_k,C_k\}$ can be seen as the number of NS5-branes in each stack (the number of NS5-branes in the $m^{th}$ stack being fixed by the condition $\sum_k \delta \eta_k = N_5$) and the values of the gauge couplings $g_{\eta_k}$ that are kept fixed, respectively.

To make the relation between $\{g_{\eta_{k}}\}$ and $\{C_k\}$ precise, we want to compute the value
of $\{g_{\eta_{k}}\}$. We can do this by considering an instanton in the relevant gauge group, whose action should be $8\pi^2 /g_{\eta_k}^2$. In our
gravity solution this maps to
 a Euclidean D0-brane stretched between the $k^{th}$ and $(k+1)^{th}$ stacks of NS5-branes, such that the D0-brane's world-line is embedded in the D4-branes' world-volume. One can show that solving the
 equations of motion for the embedding of this D0-brane in the $(\rho,\eta)$ plane implies that it wants to stay as close as
 possible to the cut at $\eta=\eta_k$, so that the world-line $\tau$ of the D0-brane may be parameterized by
\be
\tau = \lim_{\epsilon \rightarrow 0}  \left( \left\{ (\rho,\eta) | L_k \le \rho < \infty, \eta= \eta_k-\epsilon \right\} \bigcup \left\{ (\rho,\eta) | L_k \le \rho < \infty, \eta= \eta_k+\epsilon \right\} \right).
\e
In addition the D0-brane sits at some position $\chi(\tau)$ on the circle. We identify the action of the probe D0-brane
with the instanton action
$S_{D0}=8\pi^2 /g_{\eta_k}^2$.
 The total Euclidean action of the D0-brane is the sum of the DBI action and the Wess-Zumino term
\be
S_{\textrm{D0}} = S_{DBI} + S_{WZ}
= - T_0 \int d\tau e^{-\phi} \sqrt{h} - i  \mu_0 \int d\tau A_{\chi} \frac{d\chi}{d\tau}\,
\e
where $T_0^{-1}= \mu_0^{-1}=\sqrt{\alpha'}$. The induced metric on the D0-brane world-line $ds_{D0}^2 = h d\tau^2$ is derived from the 10-dimensional metric \eqref{eq:10d.metric}, and is given by (recall that $\kappa = (\alpha')^{3/2}$)
\be
\begin{split}
h= &  \kappa^{2/3} \lim_{\epsilon \rightarrow 0} \left(\left[g_{\rho\rho}+g_{\chi\chi}\left(\frac{d\chi}{d\rho} \right)^{2} \right]_{\eta=\eta_k-\epsilon}+ \left[g_{\rho\rho}+g_{\chi\chi}\left(\frac{d\chi}{d\rho} \right)^{2} \right]_{\eta=\eta_k+\epsilon} \right)\theta(\rho - L_k). \\
\end{split}
\e
The solution to the equation of motion for the transverse coordinate $\chi(\rho)$ is
\be
\frac{d\chi}{d\rho}= i (c-A_{\chi}) \left(\frac{g_{\rho\rho}}{e^{-2\phi} g_{\chi\chi}^2+(c-A_{\chi})^2g_{\chi\chi}}\right)^{1/2},
\e
where $c$ is an integration constant. The requirement $d\chi/d\rho \xrightarrow{\rho \rightarrow \infty} 0$ fixes $c=0$. Using that $\dot V|_{\eta=\eta_k\pm \epsilon}=V''|_{\eta=\eta_k\pm \epsilon}=0$ for infinitesimal $\epsilon$, after some algebra we find
\be
\begin{split}
S_{\textrm{D0}}= & - \int_0^{\infty} d\rho \lim_{\epsilon \rightarrow 0} \left( \partial_{\rho} V'|_{\eta=\eta_k-\epsilon} - \partial_{\rho} V'|_{\eta=\eta_k+\epsilon} \right) \theta(\rho - L_k) \\
= & \int_{L_k}^{\infty} d\rho (\partial_{\rho}V'_{k+1}-\partial_{\rho}V'_k)|_{\eta=\eta_k} = C_{k+1}-C_k.\\
\end{split}
\e
It then follows that
\be \label{relation}
\frac{8\pi^2}{g_{\eta_{k}}^2} = C_{k+1} -C_{k}.
\e
Recall that in the 't Hooft-like large $N$ limit described in section \ref{validity}, we scale
$V$ with $N$, such that the ranks of the gauge groups and numbers of flavors are proportional
to $N$ while the quiver diagram remains fixed. In this limit the right-hand side of \eqref{relation}
scales with $N$, so that the 't Hooft couplings $g_{\eta_k}^2 \lambda(\eta_k)$ of the gauge
groups with non-infinite couplings remain fixed, as expected.

Finding the relation between the $\{C_k\}$ (or the gauge couplings) and the $\{L_k\}$ is more complicated,
and seems to require finding explicit solutions to the Laplace equation. We leave this to future work.
Perhaps explicit solutions can be found with some smearing of the sources; in particular it would be interesting to understand if the solutions of \cite{Sfetsos:2010uq} can be understood in this way.

\subsection*{Acknowledgments}
\label{s:acks}

It is a pleasure to thank David Kutasov and Itamar Shamir for useful discussions.
This work was supported in part by the Israel--U.S.~Binational Science Foundation, by an Israel Science Foundation center for excellence grant, by the German-Israeli Foundation (GIF) for Scientific Research and Development, and by the Minerva foundation with funding from the Federal German Ministry for Education and Research.



\begin{thebibliography}{00}



\bibitem{Ganor}
O.~Ganor, unpublished.


\bibitem{Hanany:1996ie}
  A.~Hanany and E.~Witten,
  ``Type IIB superstrings, BPS monopoles, and three-dimensional gauge dynamics,''  Nucl.\ Phys.\ B {\bf 492} (1997) 152  [hep-th/9611230].  


\bibitem{Aharony:2011yc}
  O.~Aharony, L.~Berdichevsky, M.~Berkooz and I.~Shamir,
  ``Near-horizon solutions for D3-branes ending on 5-branes,''
  Phys.\ Rev.\ D {\bf 84} (2011) 126003
  [arXiv:1106.1870 [hep-th]].

\bibitem{Assel:2011xz}
  B.~Assel, C.~Bachas, J.~Estes and J.~Gomis,
  ``Holographic Duals of D=3 N=4 Superconformal Field Theories,''
  JHEP {\bf 1108} (2011) 087
  [arXiv:1106.4253 [hep-th]].



\bibitem{D'Hoker:2007xy}
  E.~D'Hoker, J.~Estes and M.~Gutperle,
  ``Exact half-BPS Type IIB interface solutions. I. Local solution and supersymmetric Janus,''
  JHEP {\bf 0706} (2007) 021
  [arXiv:0705.0022 [hep-th]].


\bibitem{D'Hoker:2007xz}
  E.~D'Hoker, J.~Estes and M.~Gutperle,
  ``Exact half-BPS Type IIB interface solutions. II. Flux solutions and multi-Janus,''
  JHEP {\bf 0706} (2007) 022
  [arXiv:0705.0024 [hep-th]].



\bibitem{Gaiotto:2008ak}
  D.~Gaiotto and E.~Witten,
  ``S-Duality of Boundary Conditions In N=4 Super Yang-Mills Theory,''
  arXiv:0807.3720 [hep-th].


\bibitem{Witten:1997sc}
  E.~Witten,
  ``Solutions of four-dimensional field theories via M theory,''  Nucl.\ Phys.\ B {\bf 500} (1997) 3  [hep-th/9703166].  


\bibitem{Maldacena:1997re}
  J.~M.~Maldacena,
  ``The Large N limit of superconformal field theories and supergravity,''  Adv.\ Theor.\ Math.\ Phys.\  {\bf 2} (1998) 231   [Int.\ J.\ Theor.\ Phys.\  {\bf 38} (1999) 1113]  [hep-th/9711200].  


\bibitem{Gubser:1998bc}
  S.~S.~Gubser, I.~R.~Klebanov and A.~M.~Polyakov,
  ``Gauge theory correlators from noncritical string theory,''  Phys.\ Lett.\ B {\bf 428} (1998) 105  [hep-th/9802109].  


\bibitem{Witten:1998qj}
  E.~Witten,
  ``Anti-de Sitter space and holography,''  Adv.\ Theor.\ Math.\ Phys.\  {\bf 2} (1998) 253  [hep-th/9802150].  




\bibitem{Lin:2004nb}
  H.~Lin, O.~Lunin and J.~M.~Maldacena,
  ``Bubbling AdS space and 1/2 BPS geometries,''  JHEP {\bf 0410} (2004) 025  [hep-th/0409174].  

\bibitem{OColgain:2010ev}
  E.~O Colgain, J.~-B.~Wu and H.~Yavartanoo,
  ``On the generality of the LLM geometries in M-theory,''
  JHEP {\bf 1104} (2011) 002
  [arXiv:1010.5982 [hep-th]].

\bibitem{Gaiotto:2009we}
  D.~Gaiotto,
  ``N=2 dualities,''  arXiv:0904.2715 [hep-th].  



\bibitem{Gaiotto:2009gz}
  D.~Gaiotto, J.~Maldacena,
  ``The Gravity duals of N=2 superconformal field theories,''
  [arXiv:0904.4466 [hep-th]].

\bibitem{Colgain:2011hb}
  E.~OColgain and B.~Stefanski, Jr.,
  ``A search for AdS5 X S2 IIB supergravity solutions dual to N = 2 SCFTs,''
  JHEP {\bf 1110} (2011) 061
  [arXiv:1107.5763 [hep-th]].

\bibitem{ReidEdwards:2010qs}
  R.~A.~Reid-Edwards, B.~Stefanski, jr.,
  ``On Type IIA geometries dual to N = 2 SCFTs,''
  Nucl.\ Phys.\  {\bf B849 } (2011)  549-572.
  [arXiv:1011.0216 [hep-th]].



\bibitem{Gadde:2009dj}
  A.~Gadde, E.~Pomoni and L.~Rastelli,
  ``The Veneziano Limit of $\mathcal N=2$ Superconformal QCD: Towards the String Dual of $\mathcal N=2$ $SU(N_c)$ SYM with $N_f = 2 N_c$,''  arXiv:0912.4918 [hep-th].  



\bibitem{Gadde:2010zi}
  A.~Gadde, E.~Pomoni and L.~Rastelli,
  ``Spin Chains in $\mathcal N=2$ Superconformal Theories: From the $\mathbb{Z}_2$ Quiver to Superconformal QCD,''  arXiv:1006.0015 [hep-th].  

\bibitem{Passerini:2011fe}
  F.~Passerini and K.~Zarembo,
  ``Wilson Loops in N=2 Super-Yang-Mills from Matrix Model,''
  JHEP {\bf 1109} (2011) 102
   [Erratum-ibid.\  {\bf 1110} (2011) 065]
  [arXiv:1106.5763 [hep-th]].

\bibitem{Bourgine:2011ie}
  J.~-E.~Bourgine,
  ``A Note on the integral equation for the Wilson loop in N = 2 D=4 superconformal Yang-Mills theory,''
  J.\ Phys.\ A A {\bf 45} (2012) 125403
  [arXiv:1111.0384 [hep-th]].

\bibitem{Fraser:2011qa}
  B.~Fraser and S.~P.~Kumar,
  ``Large rank Wilson loops in N=2 superconformal QCD at strong coupling,''
  JHEP {\bf 1203} (2012) 077
  [arXiv:1112.5182 [hep-th]].

\bibitem{Ward:1990qt}
  R.~S.~Ward,
  ``Einstein-Weyl spaces and SU(infinity) Toda fields,''
  Class.\ Quant.\ Grav.\  {\bf 7} (1990) L95.


\bibitem{Grana:2005jc}
  M.~Grana,
  ``Flux compactifications in string theory: A Comprehensive review,''
  Phys.\ Rept.\  {\bf 423 } (2006)  91-158
  [hep-th/0509003].


\bibitem{Maldacena:2000yy}
  J.~M.~Maldacena and C.~Nunez,
  ``Towards the large N limit of pure N=1 superYang-Mills,''  Phys.\ Rev.\ Lett.\  {\bf 86} (2001) 588  [hep-th/0008001].  


\bibitem{Maldacena:2001pb}
  J.~M.~Maldacena and H.~S.~Nastase,
  ``The Supergravity dual of a theory with dynamical supersymmetry breaking,''  JHEP {\bf 0109} (2001) 024  [hep-th/0105049].  




\bibitem{Hopf:1927}
 E.~Hopf,
 ``Elementare Bemerkungen \"uber die L\"osungen partieller Differentialgleichungen zweiter
Ordnung vomelliptischen Typus,''
 Sitzungsberichte Preussiche Akademie Wissenschaften, Berlin, 1927, pp. 147–-152.


\bibitem{Marolf:2000cb}
  D.~Marolf,
  ``Chern-Simons terms and the three notions of charge,''
  arXiv:hep-th/0006117.


\bibitem{Colgain:2011hb}
  E.~OColgain and B.~Stefanski, Jr.,
  ``A search for AdS5 X S2 IIB supergravity solutions dual to N = 2 SCFTs,''
  JHEP {\bf 1110} (2011) 061
  [arXiv:1107.5763 [hep-th]].

\bibitem{Sfetsos:2010uq}
  K.~Sfetsos and D.~C.~Thompson,
  ``On non-abelian T-dual geometries with Ramond fluxes,''
  Nucl.\ Phys.\ B {\bf 846} (2011) 21
  [arXiv:1012.1320 [hep-th]].



\end{thebibliography}
\end{document}